\documentclass[%
a4paper,
superscriptaddress,
notitlepage,
 amsmath,amssymb,
 aps,
 twocolumn,
 prl
 floatfix
]{revtex4-2}

\usepackage{tikz}
\usetikzlibrary{calc,arrows,decorations.pathreplacing,backgrounds}
\colorlet{mylinkcolor}{blue!66!black!80}
\usepackage{mathtools,amssymb,amsfonts}
\usepackage[colorlinks=true,linkcolor=mylinkcolor,citecolor=mylinkcolor,filecolor=cyan,urlcolor=mylinkcolor,breaklinks=true]{hyperref}
\usepackage{xcolor}

\usepackage[utf8]{inputenc}
\usepackage{booktabs}
\usepackage{bbm,mathrsfs,mathtools}
\usepackage{graphicx}
\usepackage{hyperref}

\DeclareMathOperator{\arctanh}{arctanh}
\definecolor{Doron}{RGB}{0, 102, 204}

\begin{document}

\title{Strong Mpemba Effect Through a Reentrant Phase Transition}

\author{Kristian Blom}
\email{kristian.blom@uni-muenster.de}
\affiliation{Institute of Theoretical Physics, University of M\"unster, Wilhelm-Klemm-Str.\ 9, 48149 M\"unster, Germany}
\affiliation{Center for Data Science and Complexity (CDSC),
University of M\"{u}nster, Corrensstrasse 2,  48149 M\"{u}nster, Germany}

\author{Doron Benyamin}
\affiliation{Department of Physics of Complex Systems, Weizmann Institute of Science, 7610001 Rehovot, Israel}

\author{Uwe Thiele}
\affiliation{Institute of Theoretical Physics, University of M\"unster, Wilhelm-Klemm-Str.\ 9, 48149 M\"unster, Germany}
\affiliation{Center for Data Science and Complexity (CDSC),
University of M\"{u}nster, Corrensstrasse 2,  48149 M\"{u}nster, Germany}

\author{Oren Raz}
\affiliation{Department of Physics of Complex Systems, Weizmann Institute of Science, 7610001 Rehovot, Israel}

\author{Alja\v{z} Godec}
\email{agodec@physik.uni-freiburg.de}
\affiliation{Mathematical Physics and Stochastic Dynamics, Institute
  of Physics, University of Freiburg, 79104 Freiburg im Breisgau,
  Germany}
\affiliation{Mathematical bioPhysics Group, Max Planck Institute for Multidisciplinary Sciences, Göttingen 37077, Germany}

\begin{abstract}
We investigate temperature quenches across the reentrant phase transition of the antiferromagnetic Ising model in a magnetic field and show that it provides a natural mechanism for strong direct and inverse Mpemba effects. For quenches terminating in the paramagnetic phase, the slowest relaxation mode is purely staggered. Initial states in the paramagnetic phase therefore have exactly zero overlap with this mode and exhibit a strong Mpemba effect, whereas antiferromagnetic initial states excite it and develop a slow-relaxation tail. Moreover, reentrance makes the equilibrium staggered magnetization nonmonotonic, producing conventional direct and inverse Mpemba effects when both initial states lie in the antiferromagnetic phase and the quench terminates in the paramagnetic phase. By varying the lattice coordination number, we show that this mechanism disappears in the absence of reentrance. Our results provide the first demonstration of (strong) Mpemba effects in the antiferromagnetic Ising model within the pair approximation and establish a direct link between anomalous relaxation and equilibrium phase behavior.
\end{abstract}

\maketitle
%-------------------------------
%-------------------------------
%-------------------------------
Systems coupled to a thermal environment 
eventually attain the temperature of their
surroundings in a process called \emph{thermal relaxation}
\cite{MazurGroot}. Relaxation close to
equilibrium was explained by 
linear response theory, resting on
Onsager's regression hypothesis
\cite{onsager_1,onsager_2,Yokota}: relaxation after a weak temperature quench is
equivalent to that from a 
spontaneous thermal fluctuation at equilibrium. Far from equilibrium,
linear-response arguments fail and \emph{anomalous
relaxation phenomena} emerge, such as the Kovacs effect \cite{bertin2003kovacs,bouchbinder2010nonequilibrium,santos2024mpemba,lahini2017nonmonotonic}, the Mpemba effect \cite{PhysRevLett.134.107101,E_B_Mpemba_1969,TEZA20261,doi:10.1073/pnas.1701264114,PhysRevX.9.021060,Kumar_2022,PhysRevLett.131.017101,Deguenther2022EPL}  and
the thermal relaxation asymmetry \cite{Lapolla2020PRL,VanVu2021PRR,Manikandan2021PRR,Meibohm2021PRE,Dieball2023PRR,Ibanez2024NP,Dieball_26}.

The Mpemba effect refers to a class of phenomena 
where a system initially farther from thermal equilibrium relaxes
faster than one initiated closer to it
\cite{PhysRevLett.134.107101,E_B_Mpemba_1969,TEZA20261,doi:10.1073/pnas.1701264114,PhysRevX.9.021060,Kumar_2022,PhysRevLett.131.017101,doi:10.1073/pnas.1819803116}. 
One of these classes can be understood within Markovian dynamics in
terms of the spectral properties: %decomposition of the dynamics
initial states with a suppressed projection onto the slowest relaxation mode equilibrate faster \cite{doi:10.1073/pnas.1701264114}.
This framework unifies the standard and inverse Mpemba effects and enables quantitative characterizations, such as the strong Mpemba effect and the Mpemba index \cite{doi:10.1073/pnas.1701264114,PhysRevX.9.021060}. 
Extensions to quantum systems \cite{quantum1,
  PhysRevLett.133.010402,PhysRevB.100.125102,quantum_review,
  rbt4-psfd,PhysRevLett.133.010403, PhysRevLett.131.080402,PhysRevLett.134.220403}, and experimental demonstrations 
\cite{PhysRevE.102.012906,10.1119/1.2186331,PhysRevLett.119.148001,bechoefer,Kumar_2022}
further highlight the breadth of the phenomenon (for a recent account
see \cite{TEZA20261}). 

Among the various systems that showcase the Mpemba effect, spin systems are quite unique, since they demonstrate many types of different phenomena: Direct and inverse effects have been demonstrated in
spin-glass systems \cite{doi:10.1073/pnas.1701264114}; Mpemba effects
across first-order phase transitions have been found in the
Blume-Emery-Griffiths model \cite{zhang2022theoretical}, the spin-1
Nagle--Kardar model \cite{li2026minimal}, and the Ising model with a
staggered magnetic field \cite{yang2020non}; and numerical simulations
have revealed Mpemba effects in ferromagnetic systems
\cite{chatterjee2024mpemba,ghosh2025simulations}. Particularly relevant
for the present work is the antiferromagnetic Ising model in a uniform
magnetic field. In this system, nearest-neighbor antiferromagnetic
interactions favor staggered order, whereas the external field favors
parallel alignment, producing a competition that gives rise to a phase
boundary between antiferromagnetic and paramagnetic phases. This phase
boundary has been extensively studied using exact arguments, series
expansions, Monte Carlo simulations, and analytical approximations
\cite{10.1098/rspa.1940.0086,10.1063/1.1748495,JMZiman_1951,KASTELEIJN1956387,vives1997unified,muller1977interface,PhysRevB.21.1941,10.1098/rspa.1960.0005,10.1098/rspa.1960.0122}.
Within this antiferromagnetic setting, an Mpemba effect through a
second-order phase transition was shown in Landau theory
\cite{Holtzman}, while final-relaxation Mpemba effects of various kinds
were demonstrated in mean-field \cite{PhysRevX.9.021060}, one-dimensional
\cite{PhysRevLett.131.017101,doi:10.1073/pnas.1701264114}, and
two-dimensional \cite{raz2020pre} antiferromagnetic systems.

The competition between staggered ordering and field-induced
alignment also makes the antiferromagnetic Ising model a natural setting
for \emph{reentrant phase behavior}, in which increasing
antiferromagnetic order upon cooling is followed by a return to disorder
at lower temperatures. Such reentrance has been observed in
magnetic compounds \cite{ARUGAKATORI19921639,experiment2,experiment_3}, spin-glass systems
\cite{PhysRevB.57.10264,PhysRevE.84.040101}, frustrated
Ising lattices \cite{T_Morita_1986,KITATANI198545}, and
superconductors \cite{PhysRevX.15.021075}. In the nearest-neighbor
antiferromagnetic Ising model on the square lattice, however, the phase
boundary remains monotonic
\cite{muller1977interface,PhysRevB.21.1941,pomares2026unravelinganomalousrelaxationeffects},
whereas approximate treatments of higher-connectivity or tree-like
lattices predict reentrant critical lines
\cite{BURLEY1961768,SHIRLEY1972183,PhysRevB.38.802,MONROE1994218,PhysRevB.70.224436,PhysRevB.73.214439}.
Together, these results suggest that lattice connectivity is a central
factor controlling whether reentrance can occur, and they motivate a
closer examination of how this equilibrium feature shapes relaxation
phenomena such as the Mpemba effect.

Here we show that reentrant phase transitions provide a natural mechanism for anomalous relaxation in the antiferromagnetic Ising model.~Using the Bethe–Guggenheim (pair) approximation \cite{JMZiman_1951,KASTELEIJN1956387,10.1143/PTP.51.82,BURLEY1961768,huckaby2005simple,blom2023thermodynamically,blom2023pair}, we analyze single spin-flip Glauber dynamics on lattices with varying coordination number and demonstrate the emergence of the strong (inverse) Mpemba effect for temperature quenches through a reentrant phase transition. The reentrant phase diagram allows two quenches in the same
temperature direction toward a paramagnetic 
state, one starting
    in the paramagnetic 
    and the other in the antiferromagnetic phase.  As a
result, their coupling to the slowest mode is qualitatively different. 
By varying the coordination number we show that the strong Mpemba effects disappears when the phase diagram becomes monotonic and reentrance is absent. 
Our results therefore establish a direct connection between anomalous relaxation and the equilibrium phase diagram.
%-------------------------------
%-------------------------------
%-------------------------------

\textit{Lattice setup and dynamics.}---We consider the Ising model on a Bethe lattice with coordination number $z$ in the presence of a uniform magnetic field, described by the Hamiltonian
\begin{equation}
    \mathcal{H}(\boldsymbol{\sigma}) = -J\sum_{\langle ij\rangle}\sigma_{i}\sigma_{j} - H \sum_{i} \sigma_{i},
    \label{H}
\end{equation}
where $\boldsymbol{\sigma}=\{\sigma_{1},\dots,\sigma_{N}\}$ with
$\sigma_i=\pm1$, $i\in\{1,\dots,N\}$ denotes a spin configuration,
${J<0}$ is the antiferromagnetic nearest-neighbor coupling constant,
$H$
the magnetic field, and $\langle ij \rangle$ denotes nearest-neighbor pairs.  

The dynamics is governed by single spin-flip Glauber dynamics \cite{glauber_timedependent_1963}. Let $P(\boldsymbol{\sigma};t)$ denote the probability for configuration $\boldsymbol{\sigma}$ at time $t$, which obeys the master equation
${\rm d}P(\boldsymbol{\sigma};t)/{\rm d}t
{=}\sum_{i}[
w(\boldsymbol{\sigma}^{\prime}_{i}{\rightarrow}\boldsymbol{\sigma})P(\boldsymbol{\sigma}^{\prime}_{i};t)
{-}
w(\boldsymbol{\sigma}{\rightarrow} \boldsymbol{\sigma}^{\prime}_{i})P(\boldsymbol{\sigma};t)
]$
where $\boldsymbol{\sigma}^{\prime}_{i}$ is obtained from $\boldsymbol{\sigma}$ by flipping spin $i$. The transition rates satisfy detailed balance with respect to the Boltzmann distribution associated with $\mathcal{H}(\boldsymbol{\sigma})$. Assuming nearest-neighbor interactions and spatial isotropy, this gives $w(\boldsymbol{\sigma}\rightarrow \boldsymbol{\sigma}^{\prime}_{i})
    =
    \left[1-\tanh{\left(\beta[\mathcal{H}(\boldsymbol{\sigma}^{\prime}_{i})-\mathcal{H}(\boldsymbol{\sigma})]/2\right)}\right]/2\tau$,
where $\tau$ is an intrinsic timescale for a spin-flip attempt, and $\beta \equiv 1/k_{\rm B}T$ is the inverse thermal energy \cite{glauber_timedependent_1963}. 
%-------------------------------
%-------------------------------

\textit{Order parameters.}---To capture antiferromagnetic order, we
partition the Bethe lattice into two interpenetrating sublattices
\cite{vives1997unified}, denoted by $\Lambda^a$ and $\Lambda^b$ with
cardinality $|\Lambda^i|$, such
that nearest neighbors of a site in $\Lambda^a$ belong to $\Lambda^b$
and vice versa. This bipartite structure naturally reflects the Néel
ordering tendency for $J<0$. We define the sublattice magnetizations
\begin{equation}
    m^{\mu}(t)\equiv \frac{1}{|\Lambda^{\mu}|}\sum_{i\in \Lambda^{\mu}}\langle \sigma_i \rangle(t), 
    \qquad \mu \in\{a,b\},
\end{equation}
where $\langle f \rangle(t) \equiv \sum_{\boldsymbol{\sigma}}P(\boldsymbol{\sigma};t)f(\boldsymbol{\sigma})$. From these we construct the total magnetization $m(t)$ and staggered magnetization $s(t)$
\begin{align}
    \begin{split}
    m(t)&\equiv [m^a(t)+m^b(t)]/2\in[-1,1], \\
    s(t)&\equiv [m^a(t)-m^b(t)]/2\in[-1,1],
    \end{split}
    \label{msdef}
\end{align}
where we have assumed that $|\Lambda^{a}|=|\Lambda^{b}|$.
The total magnetization $m(t)$ characterizes ferromagnetic alignment, while the staggered magnetization $s(t)$ is the order parameter for antiferromagnetic (Néel) order and measures spontaneous sublattice symmetry breaking. In addition, we introduce the nearest-neighbor correlation
\begin{equation}
    q(t) \equiv \frac{2}{zN}\sum_{\langle ij \rangle} \langle \sigma_i \sigma_j\rangle(t) \in [-1,1],
\end{equation}
which quantifies local alignment between neighboring spins and is directly related to the interaction energy in Eq.~\eqref{H}. In the following, we use the notation $\boldsymbol{x}\equiv (m,s,q)$ to denote the set of macroscopic variables. 
%-------------------------------
%-------------------------------

\textit{Evolution equations.}---Within the BG approximation \cite{PhysRevE.111.024207}, the spin-flip dynamics can be reduced to a closed set of equations for $\boldsymbol{x}_t\equiv \boldsymbol{x}(t)$ \cite{Kubo}. Our first main result is an expression for the coarse-grained transition rates to flip a spin in state $\sigma=\pm$ on sublattice $\mu\in\{a,b\}$ with $l\in\{0,1,...,z\}$ surrounding up neighbors \footnote{See Supplemental Material at...}
\begin{align}
w_{l}^{\mu \pm}(\boldsymbol{x}_t;T) \equiv \mathcal{P}^{\mu \pm}_{l}(\boldsymbol{x}_t)[1{\mp} \tanh([2l{-}z]\tilde{J} {+} \tilde{H})]/2\tau,
\end{align}
where $\mathcal{P}^{\mu\pm}_l(\boldsymbol{x})\in[0,1]$ denotes the probability that the central spin on sublattice $\mu$ has this specific local environment of neighboring spins in the macrostate $\boldsymbol{x}$, and $(\tilde{J},\tilde{H})=\beta (J,H)$. In the thermodynamic limit $N\rightarrow \infty$ this probability reads
\begin{align}
\mathcal{P}^{\mu\pm }_{l}(\boldsymbol{x})
= \binom{z}{l}
\frac{(1\pm m^{\mu})^{1-z}}{2^{z+1}}
\frac{(1\pm m^{\mu}{+}m^{\nu} \pm q)^{l}}
     {(1\pm m^{\mu}{-}m^{\nu}\mp q)^{l-z}},
\label{Pbg}
\end{align}
with $\nu\neq\mu$. In terms of the coarse-grained transition rates, the BG evolution equations take the form \cite{Note1}
\begin{align}
\begin{split}
 \frac{{\rm d} m^{\mu}(t)}{{\rm d}t}
&= 2\sum_{l=0}^{z} [w_{l}^{\mu -}(\boldsymbol{x}_t;T)-w_{l}^{\mu +}(\boldsymbol{x}_t;T)], \\
\frac{{\rm d} q(t)}{{\rm d}t}
&= 2\sum_{\mu}\sum_{l=0}^{z}
\alpha_l[w_{l}^{\mu -}(\boldsymbol{x}_t;T)-w_{l}^{\mu +}(\boldsymbol{x}_t;T)],
\end{split}
\label{BGeqs}
\end{align}
with $\alpha_l\equiv 2l/z-1$.
Equation~\eqref{BGeqs}, together with Eq.~\eqref{Pbg}, constitutes a closed set of three autonomous first-order differential equations for $(m^a,m^b,q)$, and thus for $\boldsymbol{x}$ via Eq.~\eqref{msdef}. Equilibrium states correspond to stationary solutions of Eq.~\eqref{BGeqs}, which we analyze in the next section.
\begin{figure*}[t!]
    \centering
    \includegraphics[width=\linewidth]{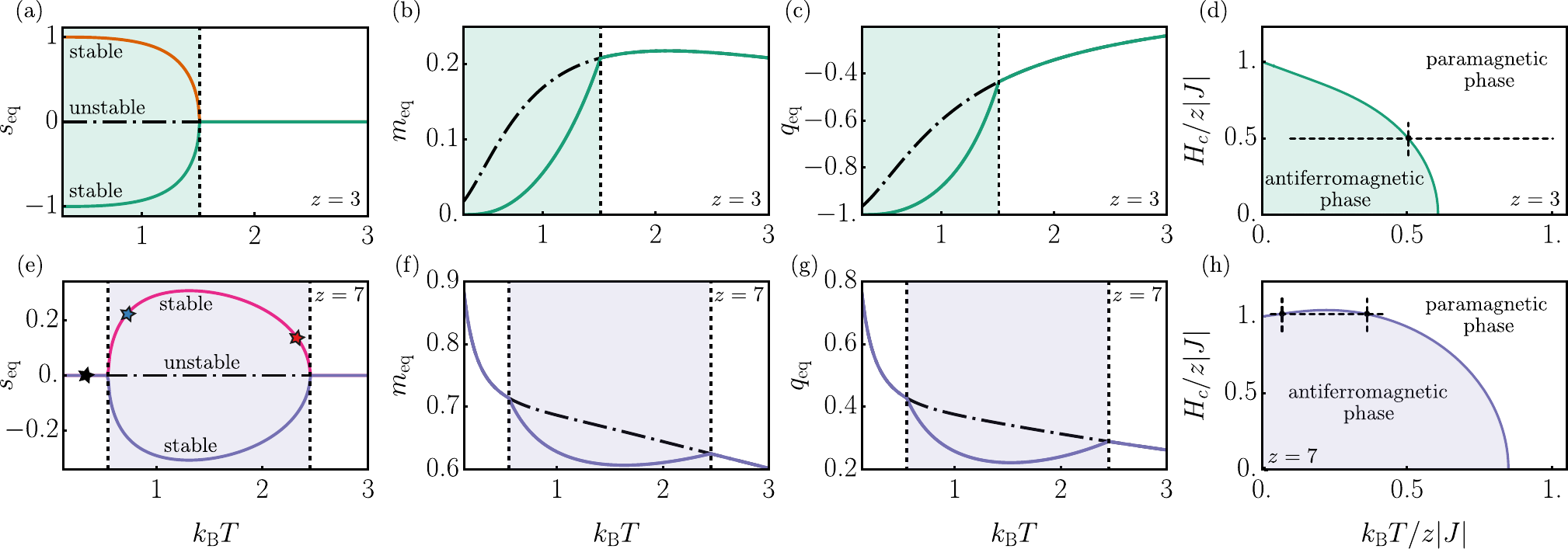}
\caption{\textbf{Emergence of reentrant phase behavior with increasing coordination number.}
(a)–(c) Equilibrium values [Eq.~\eqref{min}] of the staggered magnetization $s$, total magnetization $m$, and nearest-neighbor correlation $q$ as functions of temperature $T$ at fixed magnetic field $H/z|J|=0.5$ for $z=3<z^{*}\approx5.14$. Solid lines denote stable equilibria, dashed lines unstable ones. (d) Corresponding phase diagram in the $(H,T)$ plane. The critical line (green) is monotonic and no reentrant phase transition is observed.
(e)–(g) Equilibrium values of $(s,m,q)$ for $z=7>z^{*}$ and $H/z|J|=1.02$. In contrast to $z=3$, the staggered magnetization is nonzero only within an intermediate temperature window. In panel (e), the blue and red stars mark two antiferromagnetic initial states, while the black star marks a low-temperature paramagnetic final state (see further explanation in main text). (h) Corresponding phase diagram, where the critical line (purple) becomes nonmonotonic and a reentrant phase transition emerges at low temperature. In all panels $J{=}-1$.}
    \label{Fig1}
\end{figure*}
%-------------------------------
%-------------------------------

\textit{Equilibrium.}---The stationary solutions of Eq.~\eqref{BGeqs} coincide with the minima of the BG free energy, whose derivation is our second result and can be written as \cite{Note1}
\begin{equation}
    \mathcal{F}(\boldsymbol{x};T)
    =
    -zJq/2
    - H m
    - \beta^{-1}\mathcal{S}_{\rm BG}(\boldsymbol{x}).
    \label{FBG}
\end{equation}
The first two terms represent the energetic contributions from nearest-neighbor interactions and the external field, while the remaining terms encode entropic contributions arising from the multiplicity of spin configurations which is explicitly given in \cite{Note1}. As shown in Appendix~A, $\mathcal{F}(\boldsymbol{x};T)$ is a Lyapunov function of the dynamics~\eqref{BGeqs}, i.e., ${\rm d}\mathcal{F}(\boldsymbol{x}_t;T)/{\rm d}t\leq 0$. Consequently, equilibrium steady states correspond to the minima of the BG free energy,
\begin{equation}
\boldsymbol{x}_{\rm eq}(T)\equiv (m_{\rm eq},s_{\rm eq},q_{\rm eq})
=\arg\min\nolimits_{\boldsymbol{x}} \mathcal{F}(\boldsymbol{x};T),
\label{min}
\end{equation}
which can be determined numerically or, in certain limits, analytically \cite{PhysRevX.11.031067}. The resulting equilibrium values are shown in Fig.~\ref{Fig1} as a function of temperature $T$ for $z=3$ (a–c) and $z=7$ (e–g).
For $J<0$ two equilibrium phases arise. For sufficiently small $H$ and $T$, the system resides in an antiferromagnetic phase characterized by $s_{\rm eq}\neq 0$, signaling Néel order. In contrast, at large $H$ or high $T$, the system is paramagnetic with $s_{\rm eq}=0$. The phase boundary between antiferromagnetic and paramagnetic order is determined by the loss of stability of the stationary solution at $s=0$. Within the BG approximation this yields the following analytical expression for the critical magnetic field
\begin{figure*}[t!]
    \centering
    \includegraphics[width=\linewidth]{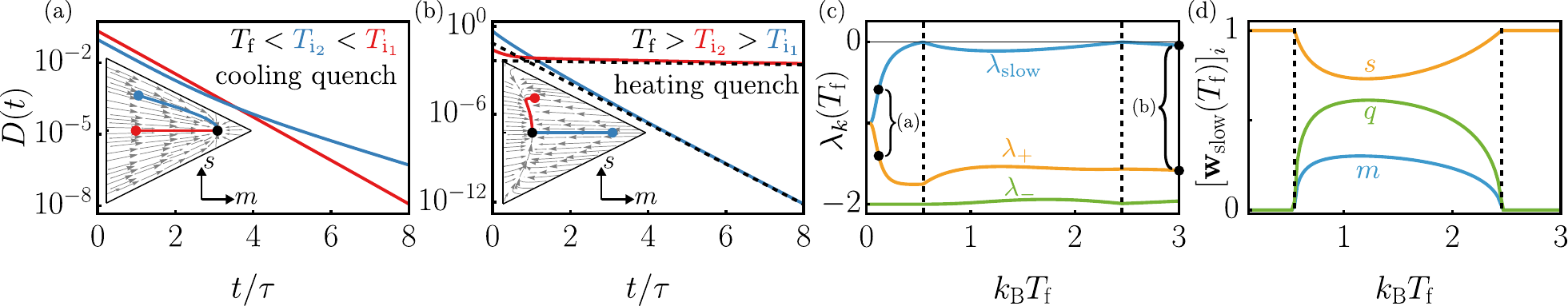}
 \caption{\textbf{Mpemba and inverse Mpemba effects across the reentrant regime and their spectral origin.}
(a) Displacement from equilibrium $D(t)$ [Eq.~\eqref{Dt}] for a cooling quench from $T_{{\rm i}_1}=3$ (paramagnetic phase) and $T_{{\rm i}_2}=2$ (antiferromagnetic phase) to $T_{\rm f}=0.12$ (paramagnetic phase). Although the quench from $T_{{\rm i}_2}$ starts closer to equilibrium, the relaxation curves cross at finite time, showing the Mpemba effect. Insets show the corresponding relaxation trajectories in the $(m,s)$ plane, zoomed in around the trajectories; blue and red dots denote the initial states, and the black dot the final equilibrium.
(b) Same protocol but with $T_{\rm f}$ and $T_{{\rm i}_2}$ interchanged, corresponding to a heating quench. The relaxation curves again cross, demonstrating the inverse Mpemba effect. Insets as in (a). The black dashed lines show the asymptotic late-time scaling of the excess free energy. Analytical expressions for the upper and lower dashed lines are given by Eqs.~\eqref{Dslow} and \eqref{Dfast}, respectively.
(c) Eigenvalues $\lambda_{k}$ as a function of the final temperature $T_{\rm f}$. The two vertical black dashed lines mark the temperature interval in which the equilibrium state is antiferromagnetic. The black points mark the values of the slowest and second-slowest eigenvalues for the two quench temperatures used in panels (a) and (b). Analytical expressions for the eigenvalues in the paramagnetic phase are given by Eqs.~\eqref{lambdaslow} and \eqref{lambdapm}. 
(d) Components of the slowest left eigenvector $\mathbf{w}_{\rm slow}$ as a function of the final temperature $T_{\rm f}$. Within the paramagnetic phase, the slow mode points purely along the staggered direction, i.e., $\mathbf{w}_{\rm slow}=(0,1,0)^{\rm T}$. In all panels we set $z=7$, $H/z|J|=1.02$, and $J=-1$.}
    \label{Fig2}
\end{figure*}
$\tilde{H}_{c}(T)
    =
    \arctanh[y]-(z-1)\arctanh[\tanh(\tilde{J})y]$,
where $y\equiv ([z-1+1/\tanh(\tilde{J})]/[z-1+\tanh(\tilde{J})])^{1/2}$, 
which is consistent with earlier analyses
\cite{huckaby2005simple,KASTELEIJN1956387,10.1143/PTP.51.82}.  The
critical magnetic field is shown in Fig.~\ref{Fig1}d,h for $z=3$ and
$z=7$, demonstrating that increasing lattice connectivity drives the
system from a conventional monotonic phase boundary (d) to one
exhibiting a nonmonotonic temperature dependence (h). Due to the
increase of 
$\tilde{H}_{c}$ at low temperature, there exists a
range of values of $H$ and $T$ for which the antiferromagnetic phase
appears upon cooling and disappears again upon further cooling, a
phenomenon known as a reentrant phase transition. To better understand
this
reentrant
behavior,
we consider the limit $T\to 0$ ($\beta\to\infty$), which yields
\begin{equation}
    \!H_{c}(T)
    {\simeq}
    {-}zJ
    {+} (k_{\rm B}T/2)
    \left[
        z\ln(z{-}2)
        {-} (z{-}1)\ln(z{-}1)
    \right],
\end{equation}
implying that slope of the phase boundary at low temperatures depends sensitively on the coordination number. 
In particular, there exists a critical value $z^{*}\approx 5.14$ determined by the transcendental equation
$(z^{*}-2)^{z^{*}} = (z^{*}-1)^{z^{*}-1}$,
at which the slope of $H_c(T)$ changes sign. For $z<z^{*}$, the critical field
decreases with increasing temperature at low $T$, yielding a monotonic
phase boundary and conventional phase behavior, as illustrated in
Fig.~\ref{Fig1}b–d. This is consistent with the exact result for the
square lattice \cite{muller1977interface}. In contrast, for $z>z^{*}$
the low-temperature slope becomes positive and the critical line turns
non-monotonic as shown in Fig.~\ref{Fig1}f–h \cite{Note1}. Such reentrant behavior is in
qualitative agreement with results reported for the body-centered
cubic lattice ($z=8$)
\cite{BURLEY1961768,SHIRLEY1972183,PhysRevB.38.802,MONROE1994218,PhysRevB.70.224436}. The
BG approximation therefore provides a unified framework in which
reentrance emerges naturally with increasing lattice connectivity. By
contrast, the mean-field approximation predicts a
reentrant phase transition for all $z$ \cite{vives1997unified}, and
thus fails to reproduce the correct square-lattice behavior. 
%-------------------------------
%-------------------------------

\textit{Temperature quenches through the reentrant regime.}---We now investigate nonequilibrium relaxation following temperature quenches across the reentrant segment of the phase diagram. 
We fix the magnetic field $H$ such that, for $z>z^{*}$, a horizontal line at this field intersects the nonmonotonic critical curve $H_c(T)$ twice (see Fig.~\ref{Fig1}h, black dashed line). 
In this regime, the system is paramagnetic at sufficiently high and sufficiently low temperatures, while antiferromagnetic order exists only within an intermediate temperature window. We will only focus on quenches whose final temperature $T_{\rm f}$ lies in the paramagnetic phase \footnote{Quenches terminating in the antiferromagnetic phase are not considered, because if the dynamics starts with $s(0)=0$ it remains trapped in this metastable state and therefore cannot relax toward the true equilibrium state.}. We consider two quench protocols that share the same final temperature $T_{\rm f}$ but originate from different initial temperatures $T_{{\rm i}_1}\neq T_{{\rm i}_2}$.  In each protocol the system is prepared in thermal equilibrium at $T_{\rm i}$ and, at time $t=0$, the temperature is instantaneously switched to $T_{\rm f}$. Relaxation toward the final equilibrium state $\boldsymbol{x}_{\rm eq}$ is quantified by the excess free energy \cite{PhysRevX.9.021060,PhysRevLett.125.110602}
\begin{equation}
    D(t)=\mathcal{F}(\boldsymbol{x}_t;T_{\rm f})
    -\mathcal{F}(\boldsymbol{x}_{\rm eq}(T_{\rm f});T_{\rm f}),
    \label{Dt}
\end{equation}
which coincides with the Kullback-Leibler divergence per spin between the time-dependent probability distribution and the final equilibrium distribution in the thermodynamic limit (see \cite{Note1}).
Since $\mathcal{F}(\boldsymbol{x};T)$ is a Lyapunov function of the
dynamics, $D(t)$ decreases monotonically and satisfies
$\lim_{t\to\infty} D(t)=0$ \footnote{The excess free energy is not
a metric in the strict
mathematical sense, as it does not
satisfy the triangle inequality \cite{Ibanez2024NP}.}. Figure~\ref{Fig2}a displays $D(t)$ for two cooling quenches with $T_{\rm f}<T_{{\rm i}_2}<T_{{\rm i}_1}$. Here $T_{{\rm i}_2}$ lies within the antiferromagnetic phase, whereas both $T_{\rm f}$ and $T_{{\rm i}_1}$ are in the paramagnetic phase. 
Because $T_{{\rm i}_2}$ is closer to $T_{\rm f}$, the corresponding protocol initially satisfies $D_2(0)<D_1(0)$, which we prove in Appendix B.  Nevertheless, the relaxation curves cross at finite time and $D_1(t)<D_2(t)$ thereafter, indicating that the system prepared at the higher initial temperature relaxes faster. 
This constitutes the Mpemba effect for quenches across the reentrant transition \cite{PhysRevX.9.021060}.  Figure~\ref{Fig2}b shows the reverse heating protocol obtained by interchanging $T_{\rm f}$ and $T_{{\rm i}_1}$. 
Although again $D_2(0){<}D_1(0)$, the curves cross at later times,
demonstrating the inverse Mpemba effect \cite{PhysRevX.9.021060}. 
%-------------------------------
%-------------------------------

\textit{Spectral decomposition.---}To elucidate the origin of these crossings, we analyze the spectral structure of the dynamics at the final temperature $T_{\rm f}$. Linearizing Eq.~\eqref{BGeqs} around the stationary point $\boldsymbol{x}_{\rm eq}(T_{\rm f})$ yields ${\rm d}\boldsymbol{\delta x}_t/{\rm d}t = \mathbf{M}\boldsymbol{\delta x}_t$, where $\boldsymbol{\delta x}_t=(\delta m(t),\delta s(t),\delta q(t))^{\rm T}$ and $\mathbf{M}$ is the Jacobian evaluated at $T_{\rm f}$. The solution can be decomposed into relaxation modes \cite{PhysRevX.9.021060},
\begin{equation}
\boldsymbol{\delta x}_t = \sum\nolimits_{k} a_k e^{\lambda_k t}\mathbf{v}_k,
\label{specdecomposition}
\end{equation}
where $\lambda_k(T_{\rm f})$ and $\mathbf{v}_k(T_{\rm f})$ denote the eigenvalues and right eigenvectors, of $\mathbf{M}$, respectively. Explicit expressions for the eigenvalues and eigenvectors are given in Appendix~D, where we also show that $\mathbf{M}$ is similar to a symmetric matrix and therefore has a real spectrum. The temperature dependence of the three eigenvalues is shown in Fig.~\ref{Fig2}c. Under a timescale separation the long-time dynamics is governed by the slowest eigenvalue $\lambda_{\rm slow}$, with right and left eigenvectors denoted by $\mathbf{v}_{\rm slow}$ and $\mathbf{w}_{\rm slow}$. Choosing the left and right eigenvectors biorthonormal, $\mathbf{w}_k^{\rm T}\mathbf{v}_{k'}=\delta_{kk'}$, the associated amplitude onto the slowest eigenmode is given by
\begin{equation}
a_{\rm slow}(T_{\rm i},T_{\rm f}) = \mathbf{w}^{\rm T}_{\rm slow}(T_{\rm f}) [ \boldsymbol{x}_{\rm eq}(T_{\rm i}) - \boldsymbol{x}_{\rm eq}(T_{\rm f}) ].
\label{aslow}
\end{equation}
For $T_{\rm f}$ in the paramagnetic phase, the slowest mode points
purely along the staggered direction, $\mathbf{v}_{\rm
  slow}=\mathbf{w}_{\rm slow}=(0,1,0)$ in the $(m,s,q)$ basis, as
shown in Fig.~\ref{Fig2}d. The corresponding overlap therefore probes
only the staggered component of the initial state. As a result,
quenches starting in the paramagnetic phase, where $s_{\rm eq}(T_{\rm
  i})=0$, have vanishing slow-mode amplitude, $a_{\rm slow}(T_{\rm
  i},T_{\rm f})=0$, corresponding to the \emph{strong Mpemba effect}
\cite{PhysRevX.9.021060}. By contrast, quenches from the
antiferromagnetic phase have $s_{\rm eq}(T_{\rm i})\neq 0$ and thus
excite the slow mode, yielding $a_{\rm slow}(T_{\rm i},T_{\rm f})=
s_{\rm eq}(T_{\rm i})$. This selective excitation of the staggered
relaxation mode is the origin of both the Mpemba and inverse Mpemba
effects. As evidenced by the \emph{qualitatively different} relaxation trajectories in the insets of Fig.~\ref{Fig2}a,b, the state with nonzero staggered magnetization relaxes with a finite projection onto the slow staggered mode. As a result, one trajectory exhibits the slow exponential decay associated with
$\lambda_{\rm slow}$, while the other follows the faster asymptotic
decay shown by the dashed lines in Fig.~\ref{Fig2}b. Hence, although the former may initially lie
closer to equilibrium, it is eventually overtaken by the state that
does not excite the slow mode, producing a crossing in
$D(t)$. In \cite{Note1}, we further show that both
the strong Mpemba and strong inverse Mpemba effects satisfy the
thermomajorization criterion and therefore hold for any distance
measure that decreases monotonically under the relaxation dynamics.

%-------------------------------
%-------------------------------
\textit{Antiferromagnetic-to-paramagnetic quenches.}--- Having identified the staggered magnetization as the slow mode for quenches ending in the paramagnetic phase, we can also conclude that reentrance guarantees Mpemba effects when both initial states lie in the antiferromagnetic phase. Indeed, reentrance makes $s_{\rm eq}(T)$ nonmonotonic, as it vanishes at both boundaries of the antiferromagnetic interval while remaining nonzero inside. For a low-temperature paramagnetic final state, we can therefore choose $T_{\rm f}<T_{{\rm i}_1}<T_{{\rm i}_2}$ such that $s_{\rm eq}(T_{{\rm i}_1})>s_{\rm eq}(T_{{\rm i}_2})$, as illustrated by the black, blue, and red stars in Fig.~\ref{Fig1}e, respectively. Although $T_{{\rm i}_2}$ (red star) is farther from equilibrium, it has the smaller slow-mode amplitude. Since $D(0)$ increases monotonically with $|T_{\rm i}-T_{\rm f}|$, the trajectory starting at $T_{{\rm i}_2}$ must eventually overtake the one starting at $T_{{\rm i}_1}$, yielding a direct Mpemba effect (see also \cite{Note1}). An analogous choice on the low-temperature side of the antiferromagnetic interval yields an inverse Mpemba effect for a high-temperature paramagnetic final state. By contrast, for $z<z^*$, where reentrance is absent, we find that $s_{\rm eq}(T)$ decreases monotonically with $T$ throughout the antiferromagnetic phase (see Fig.~\ref{Fig1}a), ruling out Mpemba effects for this type of quench.

%-------------------------------
%-------------------------------
%-------------------------------
%-------------------------------
%-------------------------------

\textit{Conclusion.}---We have shown that reentrant phase transitions provide a natural mechanism for anomalous relaxation in interacting spin systems. Using the BG description of Glauber dynamics, we demonstrated that the Mpemba and inverse Mpemba effects arise from the selective excitation of the staggered relaxation mode when quenches terminate in the paramagnetic phase. The reentrant topology enables quenches between the high- and low-temperature paramagnetic phases that do not excite the slowest mode, whereas initial states in the intermediate antiferromagnetic phase have a finite projection onto this mode. Moreover, since reentrance results in a nonmonotonic equilibrium staggered magnetization with temperature, it also gives rise to Mpemba effects when both initial states lie in the antiferromagnetic phase and the quench terminates in the paramagnetic phase. Reentrance is therefore a sufficient, but not necessary, condition for (strong) Mpemba effects. Interesting directions for future work include understanding fluctuations around the average order parameters and developing a spatially extended dynamical field theory that incorporates the staggered magnetization as an independent slow field~\cite{10.21468/SciPostPhys.20.1.005, Penrose1991,
  Ducastelle1996, blom2023thermodynamically}.

%---------------------------------------------------
%---------------------------------------------------
%---------------------------------------------------

\textit{Acknowledgements.}---We acknowledge fruitful discussions
with Marija Vucelja. Financial support from the Ministry of Research and Culture in Lower Saxony and the Volkswagen Foundation via the project \emph{Far-from-equilibrium relaxation in strongly interacting many-body systems in the presence of conservation laws} within
the program "Research Cooperations Lower Saxony--Israel''  (to A.G. and O.R.) is gratefully acknowledged.

%---------------------------------------------------
%---------------------------------------------------
%---------------------------------------------------
\appendix
%---------------------------------------------------
%---------------------------------------------------
%---------------------------------------------------
\section*{End Matter}
%---------------------------------------------------
%---------------------------------------------------
%---------------------------------------------------

\setcounter{equation}{0}
\renewcommand{\theequation}{A\arabic{equation}}

\textit{Appendix A:~Proof of the Lyapunov property.}---Using the chain rule, the time derivative of the BG free energy given by Eq.~\eqref{FBG} is (omitting the explicit arguments of $\mathcal F=\mathcal F(\boldsymbol{x};T)$)
\begin{equation}
\frac{{\rm d}\mathcal F}{{\rm d}t}=
\frac{\partial\mathcal F}{\partial q}\frac{{\rm d}q}{{\rm d}t}
+\sum_{\mu}\frac{\partial\mathcal F}{\partial m^\mu}\frac{{\rm d}m^\mu}{{\rm d}t}
.
\end{equation}
Substituting Eq.~\eqref{BGeqs} gives
\begin{equation}
\frac{{\rm d}\mathcal F}{{\rm d}t}=
2\sum_{\mu}\sum_{l=0}^z
\left[
\frac{\partial\mathcal F}{\partial m^\mu}
+
\alpha_l\frac{\partial\mathcal F}{\partial q}
\right]
(w_l^{\mu -}-w_l^{\mu +}).
\label{dFintermediate}
\end{equation}
To simplify this expression further we note that the coarse-grained transition rates are related to the free energy via the local detailed-balance relation \cite{Note1,RevModPhys.97.015002}
\begin{equation}
\frac{\partial\mathcal F}
{\partial m^{\mu}}
+
\alpha_l
\frac{\partial\mathcal F}
{\partial q}
=-\frac{1}{4\beta}\ln\biggl(
\frac{w_l^{\mu -}}
{w_l^{\mu +}}
\biggr).
\label{DB}
\end{equation}
Inserting the r.h.s.~of Eq.~\eqref{DB} into Eq.~\eqref{dFintermediate}, the time derivative of the free energy can be rewritten as
\begin{equation}
\frac{{\rm d}\mathcal F}{{\rm d}t}=-\frac{1}{2\beta}
\sum_{\mu}\sum_{l=0}^z
\bigl(w_l^{\mu -}-w_l^{\mu +}\bigr)
\ln\biggl(\frac{w_l^{\mu -}}{w_l^{\mu +}}\biggr)\leq 0,
\end{equation}
where the last inequality follows from the fact that $\beta\geq 0$, $w_l^{\mu \pm}\geq 0$, and for any $x,y>0$ one has
$(x-y)\ln{(x/y)}\geq 0$. Hence, $\mathcal F(\boldsymbol{x};T)$ given by Eq.~\eqref{FBG} is a Lyapunov function of the BG dynamics~\eqref{BGeqs}.

%---------------------------------------------------
%---------------------------------------------------
%---------------------------------------------------
\setcounter{equation}{0}
\renewcommand{\theequation}{B\arabic{equation}}

\textit{Appendix B: Monotonicity of the excess free energy.}---Here we show that the initial excess free energy
\begin{equation}
\Delta \mathcal{F}(T_{\rm i},T_{\rm f})\equiv 
\mathcal{F}(\boldsymbol{x}_{\rm eq}(T_{\rm i});T_{\rm f})
-
\mathcal{F}(\boldsymbol{x}_{\rm eq}(T_{\rm f});T_{\rm f})
\label{DF}
\end{equation}
increases monotonically with $|T_{\rm i}-T_{\rm f}|$. By construction, $\boldsymbol{x}_{\rm eq}(T_{\rm f})$ minimizes $\mathcal{F}(\boldsymbol{x};T_{\rm f})$, so that $\Delta\mathcal{F}(T_{\rm i},T_{\rm f})\ge 0$ with equality only for $T_{\rm i}=T_{\rm f}$. To determine how $\Delta\mathcal{F}(T_{\rm i},T_{\rm f})$ varies with the initial temperature, we differentiate Eq.~\eqref{DF} with respect to $T_{\rm i}$ at fixed $T_{\rm f}$. Since only the first term depends on $T_{\rm i}$, the chain rule gives
\begin{equation}
\partial_{T_{\rm i}}
\Delta\mathcal{F}(T_{\rm i},T_{\rm f})
=
\nabla_{\boldsymbol{x}}
\mathcal{F}(\boldsymbol{x}_{\rm eq}(T_{\rm i});T_{\rm f})^{\rm T}
\frac{{\rm d}\boldsymbol{x}_{\rm eq}(T_{\rm i})}{{\rm d}T_{\rm i}},
\label{DF2}
\end{equation}
where $\nabla_{\boldsymbol{x}}\equiv(\partial_m,\partial_s,\partial_q)^{\rm T}$ with $\partial_x\equiv \partial/\partial x$, and we use the shorthand notation 
\begin{equation}
    \nabla_{\boldsymbol{x}}
\mathcal{F}(\boldsymbol{x}_{\rm eq}(T_{\rm i});T_{\rm f})\equiv\left.\nabla_{\boldsymbol{x}}
\mathcal{F}(\boldsymbol{x};T_{\rm f})\right|_{\boldsymbol{x}=\boldsymbol{x}_{\rm eq}(T_{\rm i})}.
\end{equation}
Using the decomposition $\mathcal{F}(\boldsymbol{x};T)
=\mathcal{U}(\boldsymbol{x})-T\mathcal{S}_{\rm BG}(\boldsymbol{x})$ where $\mathcal{U}(\boldsymbol{x})$ is the internal energy, 
together with the equilibrium condition $\nabla_{\boldsymbol{x}}
\mathcal{F}(\boldsymbol{x}_{\rm eq}(T_{\rm i});T_{\rm i})=0$,
we can rewrite the gradient in Eq.~\eqref{DF2} as
\begin{align}
\nabla_{\boldsymbol{x}}
\mathcal{F}(\boldsymbol{x}_{\rm eq}(T_{\rm i});T_{\rm f})
&=
\nabla_{\boldsymbol{x}}[
\mathcal{F}(\boldsymbol{x}_{\rm eq}(T_{\rm i});T_{\rm f})
-
\mathcal{F}(\boldsymbol{x}_{\rm eq}(T_{\rm i});T_{\rm i})]
\nonumber\\
&=
(T_{\rm i}-T_{\rm f})
\nabla_{\boldsymbol{x}}\mathcal{S}_{\rm BG}(\boldsymbol{x}_{\rm eq}(T_{\rm i})).
\end{align}
Substituting this into Eq.~\eqref{DF2} yields
\begin{equation}
\partial_{T_{\rm i}}
\Delta\mathcal{F}(T_{\rm i},T_{\rm f})
=
(T_{\rm i}{-}T_{\rm f})
\nabla_{\boldsymbol{x}}\mathcal{S}_{\rm BG}(\boldsymbol{x}_{\rm eq}(T_{\rm i}))^{\rm T}
\frac{{\rm d}\boldsymbol{x}_{\rm eq}(T_{\rm i})}{{\rm d}T_{\rm i}}.
\label{DF3}
\end{equation}
To proceed further, we differentiate the equilibrium condition
$\nabla_{\boldsymbol{x}}
\mathcal{F}(\boldsymbol{x}_{\rm eq}(T);T)=0$
with respect to $T$. This gives
\begin{equation}
\mathbf{H}(T)
\frac{{\rm d}\boldsymbol{x}_{\rm eq}(T)}{{\rm d}T}
+
\partial_T
\nabla_{\boldsymbol{x}}\mathcal{F}(\boldsymbol{x}_{\rm eq}(T);T)=0,
\end{equation}
where $\mathbf{H}(T)$ is the Hessian of the free energy evaluated at equilibrium,
\begin{equation}
\mathbf{H}_{ij}(T)
=
\left.
\partial_{i}\partial_{j}\mathcal{F}(\boldsymbol{x};T)
\right|_{\boldsymbol{x}=\boldsymbol{x}_{\rm eq}(T)}, \ i,j\in\{m,s,q\}.
\label{HesBG}
\end{equation}
Since $\partial_T\mathcal{F}(\boldsymbol{x};T)=-\mathcal{S}_{\rm BG}(\boldsymbol{x})$, it follows that
\begin{equation}
\partial_T
\nabla_{\boldsymbol{x}}\mathcal{F}(\boldsymbol{x}_{\rm eq}(T);T)
=
-\nabla_{\boldsymbol{x}}\mathcal{S}_{\rm BG}(\boldsymbol{x}_{\rm eq}(T)).
\end{equation}
Hence, since $\mathbf{H}(T)$ is invertible at (stable) equilibrium 
\begin{equation}
\frac{{\rm d}\boldsymbol{x}_{\rm eq}(T)}{{\rm d}T}
=
\mathbf{H}^{-1}(T)
\nabla_{\boldsymbol{x}}\mathcal{S}_{\rm BG}(\boldsymbol{x}_{\rm eq}(T)).
\label{dxdt}
\end{equation}
Substituting Eq.~\eqref{dxdt} into Eq.~\eqref{DF3}, we obtain
\begin{align}
&\partial_{T_{\rm i}}
\Delta\mathcal{F}(T_{\rm i},T_{\rm f})
=
(T_{\rm i}-T_{\rm f})\times\nonumber\\
&\nabla_{\boldsymbol{x}}\mathcal{S}_{\rm BG}(\boldsymbol{x}_{\rm eq}(T_{\rm i}))^{\rm T}
\mathbf{H}^{-1}(T_{\rm i})
\nabla_{\boldsymbol{x}}\mathcal{S}_{\rm BG}(\boldsymbol{x}_{\rm eq}(T_{\rm i})).
\label{DFfinal}
\end{align}
Along a stable equilibrium branch, the Hessian $\mathbf{H}(T_{\rm i})$ is positive definite, so the quadratic form on the right-hand side of Eq.~\eqref{DFfinal} is nonnegative. Moreover, the equilibrium order parameters $\boldsymbol{x}_{\rm eq}(T)$ remain continuous across the phase boundary (see Fig.~\ref{Fig1}), and since $\mathcal{F}(\boldsymbol{x};T)$ is continuous in both $\boldsymbol{x}$ and $T$, the excess free energy $\Delta\mathcal{F}(T_{\rm i},T_{\rm f})$ is continuous in $T_{\rm i}$. Thus, although \(\partial_{T_{\rm i}}\Delta\mathcal{F}\) may be nonanalytic at the phase boundary, $\Delta\mathcal{F}$ itself cannot jump. Combined with Eq.~\eqref{DFfinal}, this implies that $\Delta\mathcal{F}(T_{\rm i},T_{\rm f})$ is minimized at $T_{\rm i}=T_{\rm f}$ and increases monotonically with $|T_{\rm i}-T_{\rm f}|$ over the full temperature range considered.
%---------------------------------------------------
%---------------------------------------------------
%---------------------------------------------------

\setcounter{equation}{0}
\renewcommand{\theequation}{C\arabic{equation}}
\textit{Appendix C: Linearized dynamics.---}Linearizing Eq.~\eqref{BGeqs} around the equilibrium stationary state $\boldsymbol{x}_{\rm eq}(T_{\rm f})=(m_{\rm eq},s_{\rm eq},q_{\rm eq})$ and using Eq.~\eqref{DB} yields
\begin{equation}
\frac{{\rm d}\boldsymbol{\delta x}_t}{{\rm d}t}
=
\mathbf{M}\boldsymbol{\delta x}_t
=
-\beta_{\rm f}\mathbf{L}\mathbf{H}\boldsymbol{\delta x}_t,
\end{equation}
where $\beta_{\rm f}=1/k_{\rm B}T_{\rm f}$, $\mathbf{H}=\mathbf{H}(T_{\rm f})$ denotes the Hessian of the free energy evaluated at the post-quench temperature [see Eq.~\eqref{HesBG}], and $\mathbf{L}=\mathbf{L}(T_{\rm f})$ is a kinetic matrix. This factorization of the Jacobian into a kinetic matrix and the Hessian follows directly from local detailed-balance and holds for any choice of transition rates obeying Eq.~\eqref{DB}. Different rate choices only modify the numerical values entering $\mathbf{L}$. The kinetic matrix is given by
\begin{equation}
\mathbf{L}(T)=\sum_{\mu,l} w^\mu_{l,{\rm eq}}(T) \mathbf{c}_l^\mu (\mathbf{c}_l^\mu)^{\rm T},
\end{equation}
where
\begin{equation}
w_{l,{\rm eq}}^\mu(T)\equiv w_l^{\mu +}(\boldsymbol{x}_{\rm eq}(T);T)=w_l^{\mu -}(\boldsymbol{x}_{\rm eq}(T);T)
\end{equation}
and
\begin{equation}
\mathbf{c}_l^a=\sqrt{8}(1/2,1/2,\alpha_l)^{\rm T},
\ 
\mathbf{c}_l^b=\sqrt{8}(1/2,-1/2,\alpha_l)^{\rm T}.
\end{equation}
Hence, $\mathbf{L}$ is symmetric and positive semidefinite. 
%---------------------------------------------------
%---------------------------------------------------
%---------------------------------------------------

\setcounter{equation}{0}
\renewcommand{\theequation}{D\arabic{equation}}
\textit{Appendix D: Spectrum in the paramagnetic phase.---}Here and in the following, we use the shorthand notation
\begin{equation}
\partial_i\partial_j \tilde{\mathcal F}_{\rm eq}
\equiv
\left.\partial_i\partial_j \tilde{\mathcal F}(\boldsymbol{x};T_{\rm f})\right|_{\boldsymbol{x}_{\rm eq}(T_{\rm f})},
\ i,j\in\{m,s,q\},
\end{equation}
where $T_{\rm f}$ is assumed to be chosen in the paramagnetic phase with $s_{\rm eq}=0$ and $w_{l,{\rm eq}}^a=w_{l,{\rm eq}}^b$. Then, the kinetic matrix takes the explicit form
\begin{equation}
\mathbf{L}=
\begin{pmatrix}
S_0 & 0 & 2S_1\\
0 & S_0 & 0\\
2S_1 & 0 & 4S_2
\end{pmatrix},
\end{equation}
with
\begin{equation}
S_k\equiv 4\sum_{l=0}^z \alpha_l^k w_{l,{\rm eq}}^\mu.
\end{equation}
We can establish that $\mathbf{L}$ is positive definite in the paramagnetic phase by inspecting its determinant, ${\rm det}(\mathbf{L})=4S_0(S_0S_2-S_1^2)$. Since $w_{l,{\rm eq}}^\mu>0$ for all $l$, one has $S_0>0$. Moreover, because the support of the equilibrium rates contains more than one value of $\alpha_l$, the Cauchy--Schwarz inequality is strict and therefore $S_0S_2-S_1^2>0$. Hence ${\rm det}(\mathbf{L})>0$. Since $\mathbf{L}$ is symmetric and positive semidefinite, this implies that $\mathbf{L}$ is positive definite in the paramagnetic phase. As a consequence, $\mathbf{L}^{1/2}$ is invertible and the Jacobian can be rewritten as $\mathbf{L}^{-1/2}\mathbf{M}\mathbf{L}^{1/2}=-\beta_{\rm f}\mathbf{L}^{1/2}\mathbf{H}\mathbf{L}^{1/2}\equiv \mathbf{K}$.
Hence, $\mathbf{M}$ is similar to the symmetric matrix $\mathbf{K}$, and therefore all eigenvalues of $\mathbf{M}$ are real. In the paramagnetic phase, we further have $\partial_m\partial_s \tilde{\mathcal{F}}_{\rm eq}=0$ and  $\partial_q\partial_s \tilde{\mathcal{F}}_{\rm eq}=0$,
and therefore the Hessian has the form
\begin{equation}
\beta_{\rm f}\mathbf{H}=
\begin{pmatrix}
\partial^2_m \tilde{\mathcal{F}}_{\rm eq} & 0 & \partial_m\partial_q \tilde{\mathcal{F}}_{\rm eq}\\
0 & \partial^2_s \tilde{\mathcal{F}}_{\rm eq} & 0\\
\partial_m\partial_q \tilde{\mathcal{F}}_{\rm eq} & 0 & \partial^2_q \tilde{\mathcal{F}}_{\rm eq}
\end{pmatrix}.
\end{equation}
It follows that the staggered mode
\begin{equation}
    \mathbf{v}_{\rm slow}=\mathbf{w}_{\rm slow}=(0,1,0)^{\rm T},
\end{equation}
is a simultaneous right and left eigenvector of $\mathbf{H}$ and $\mathbf{L}$. The corresponding eigenvalue is
\begin{equation}
\lambda_{\rm slow}
=-S_0\partial_s^2\tilde{\mathcal F}_{\rm eq} \leq 0,
\label{lambdaslow}
\end{equation}
which vanishes at the critical magnetic-field boundary.
The remaining two eigenmodes lie in the symmetric sector and take the form
\begin{align}
\mathbf{v}_{\pm}
&=\mathcal{C}_{\pm}
\left(
\!1,0,
\frac{\lambda_{\mp}{+}2S_1\partial_m\partial_q \tilde{\mathcal F}_{\rm eq}{+}4S_2\partial_q^2 \tilde{\mathcal F}_{\rm eq}}
{S_0\partial_m\partial_q \tilde{\mathcal F}_{\rm eq}{+}2S_1\partial_q^2 \tilde{\mathcal F}_{\rm eq}}
 \! \right)^{\rm T}, \\
\mathbf{w}_{\pm}&=\mathbf{L}^{-1}\mathbf{v}_{\pm},
\end{align}
where $\mathcal{C}_{\pm}$ is chosen such that $\mathbf{w}_{\pm}^{\rm T}\mathbf{v}_{\pm}=1$. Their eigenvalues are
\begin{equation}
\!\!\!\lambda_{\pm}
{=}
{-}2\left(\!
\frac{S_0}{4}\partial_m^2 \tilde{\mathcal F}_{\rm eq}
{+}S_1\partial_m\partial_q \tilde{\mathcal F}_{\rm eq}
{+}S_2\partial_q^2 \tilde{\mathcal F}_{\rm eq}
\!\right)
\!{\pm} \sqrt{\Delta},
\label{lambdapm}
\end{equation}
with
\begin{align}
\Delta
&=
4\left(
\frac{S_0}{4}\partial_m^2 \tilde{\mathcal F}_{\rm eq}
+S_1\partial_m\partial_q \tilde{\mathcal F}_{\rm eq}
+S_2\partial_q^2 \tilde{\mathcal F}_{\rm eq}
\right)^2
\nonumber\\
&
-4(S_0S_2{-}S_1^2)
\left(
\partial_m^2 \tilde{\mathcal F}_{\rm eq}\partial_q^2 \tilde{\mathcal F}_{\rm eq}
{-}(\partial_m\partial_q \tilde{\mathcal F}_{\rm eq})^2
\right),
\end{align}
and $\Delta \geq 0$ since all eigenvalues are real. From this follows that $\lambda_+\geq \lambda_-$ as seen in Fig.~\ref{Fig2}c.
%---------------------------------------------------
%---------------------------------------------------
%---------------------------------------------------

\setcounter{equation}{0}
\renewcommand{\theequation}{E\arabic{equation}}

\textit{Appendix E: Late-time scaling of the excess free energy.}---Here we derive how the excess free energy given by Eq.~\eqref{Dt} scales at late times in terms of the eigenvalues and eigenvectors of the linearized dynamics. Since $\boldsymbol{x}_{\rm eq}(T_{\rm f})$ is a stationary point of $\mathcal{F}(\boldsymbol{x};T_{\rm f})$, the linear term vanishes and one obtains
\begin{equation}
D(t)=(1/2)\boldsymbol{\delta x}_t^{\rm T}\mathbf{H}(T_{\rm f})\boldsymbol{\delta x}_t+\mathcal{O}(|\boldsymbol{\delta x}_t|^3),
\label{E1}
\end{equation}
where $\mathbf{H}(T_{\rm f})$ is the Hessian defined in Eq.~\eqref{HesBG}. Using the spectral decomposition of the linearized dynamics given by Eq.~\eqref{specdecomposition}, Eq.~\eqref{E1} becomes
\begin{equation}
D(t)=\frac{1}{2}\sum_{k,l} a_k a_l e^{(\lambda_k+\lambda_l)t}\mathbf{v}_k^{\rm T}\mathbf{H}(T_{\rm f})\mathbf{v}_l
+\mathcal{O}(|\boldsymbol{\delta x}_t|^3).
\end{equation}
At late times, the dominant contribution is determined by the slowest mode with nonvanishing amplitude. If $a_{\rm slow}\neq 0$, one has
\begin{align}
D(t)&\overset{t\to\infty}{\simeq} (1/2)a_{\rm slow}^2e^{2\lambda_{\rm slow}t}
\mathbf{v}_{\rm slow}^{\rm T}\mathbf{H}(T_{\rm f})\mathbf{v}_{\rm slow} \nonumber \\
&\overset{\textcolor{white}{t\to\infty}}{=} (1/2)a_{\rm slow}^2\partial_s^2\mathcal{F}_{\rm eq}e^{2\lambda_{\rm slow}t}\,\overset{t\to\infty}{=}0,
\label{Dslow}
\end{align}
where for the second line we used that $\mathbf{v}_{\rm slow}=(0,1,0)^{\rm T}$. This expression is shown with the upper black dashed line in Fig.~\ref{Fig2}b. If the slow mode is not excited, i.e. $a_{\rm slow}=0$, then the leading contribution comes from the next slowest mode with nonzero overlap. In particular, if $a_{+}\neq 0$, one finds
\begin{equation}
D(t)\overset{t\to\infty}{\simeq}  (1/2)a_{+}^2e^{2\lambda_{+}t}
\mathbf{v}_{+}^{\rm T}\mathbf{H}(T_{\rm f})\mathbf{v}_{+}\,\overset{t\to\infty}{=}0,
\label{Dfast}
\end{equation}
which is shown with the lower black dashed line in Fig.~\ref{Fig2}b. Thus, the exponential decay of the excess free energy is governed by twice the real part of the slowest eigenvalue with nonzero projection onto the initial displacement.
%---------------------------------
%---------------------------------
%---------------------------------
%
%---------------------------------
%---------------------------------
\counterwithout{equation}{section}
\addtocounter{equation}{-1}

\clearpage
\newpage
\onecolumngrid
\renewcommand{\thefigure}{S\arabic{figure}}
\renewcommand{\theequation}{S\arabic{equation}}
\renewcommand{\thetable}{S\arabic{table}}
\renewcommand{\thesection}{S\arabic{section}}
\renewcommand{\thesubsection}{S\arabic{subsection}}
\setcounter{equation}{0}
\setcounter{table}{0}
\setcounter{figure}{0}
\setcounter{page}{1}
\setcounter{section}{0}
%- --------------------------------
%- --------------------------------

\begin{center}\textbf{Supplementary Material for: Strong Mpemba Effect Through A Reentrant Phase Transition}\\[0.2cm]
Kristian Blom$^{1,2}$, Doron Benyamin$^{3}$, Uwe Thiele$^{1,2}$, Oren Raz$^{3}$, and Alja\v{z} Godec$^{4,5}$\\ \vspace{0.2cm}
{\footnotesize 
\emph{$^{1}$Institute of Theoretical Physics, University of M\"unster, Wilhelm-Klemm-Str.\ 9, 48149 M\"unster, Germany}\\
\emph{$^{2}$Center for Data Science and Complexity (CDSC),
University of M\"{u}nster, Corrensstrasse 2,  48149 M\"{u}nster, Germany}\\
\emph{$^{3}$Department of Physics of Complex Systems, Weizmann Institute of Science, 7610001 Rehovot, Israel}\\
\emph{$^{4}$Mathematical Physics and Stochastic Dynamics, Institute
  of Physics, University of Freiburg, 79104 Freiburg im Breisgau,
  Germany}\\
\emph{$^{5}$Mathematical bioPhysics Group, Max Planck Institute for Multidisciplinary Sciences, Göttingen 37077, Germany}}\\[0.6cm]\end{center}

\noindent In this Supplementary Material, we provide derivations and
further details about the results shown in the main paper.
%-------------------------------
%-------------------------------
%-------------------------------
%-------------------------------
%-------------------------------
%-------------------------------
\section{Derivation of the dynamical equations}
%-------------------------------
%-------------------------------
%-------------------------------
\noindent In this section we derive the dynamical equations used in the main text. We start from microscopic single-spin-flip Glauber dynamics and then apply the Bethe-Guggenheim (BG) approximation to obtain a closed set of equations for the macroscopic variables $\boldsymbol{x}=(m,s,q)$. Let $P(\boldsymbol{\sigma};t)$ denote the probability of finding the system in the spin configuration
$\boldsymbol{\sigma}=\{\sigma_{1},...,\sigma_{N}\}$ at time $t$. This probability evolves according to the master equation
\begin{equation}
\frac{{\rm d}P(\boldsymbol{\sigma};t)}{{\rm d}t}
=\sum_{i}[
w(\boldsymbol{\sigma}^{\prime}_{i}\rightarrow\boldsymbol{\sigma})P(\boldsymbol{\sigma}^{\prime}_{i};t)
-
w(\boldsymbol{\sigma}\rightarrow \boldsymbol{\sigma}^{\prime}_{i})P(\boldsymbol{\sigma};t)
],
\label{masterequation}
\end{equation}
where
$\boldsymbol{\sigma}^{\prime}_{i}=\{\sigma_{1},...,-\sigma_{i},...,\sigma_{N}\}$
is the configuration obtained from $\boldsymbol{\sigma}$ by flipping the spin at site $i$. The transition rates satisfy detailed balance with respect to the Boltzmann distribution associated with $\mathcal{H}(\boldsymbol{\sigma})$. Assuming nearest-neighbor interactions and spatial isotropy, this gives \cite{Sglauber_timedependent_1963}
\begin{equation}
    w(\boldsymbol{\sigma}\rightarrow \boldsymbol{\sigma}^{\prime}_{i})
    =
    \left[1-\tanh{\left(\beta\Delta\mathcal{H}(\boldsymbol{\sigma}\rightarrow \boldsymbol{\sigma}^{\prime}_{i})/2\right)}\right]/2\tau,
\end{equation}
where $\tau$ is an intrinsic timescale for a spin-flip attempt, $\beta \equiv 1/k_{\rm B}T$ is the inverse thermal energy, and $\Delta\mathcal{H}(\boldsymbol{\sigma}\rightarrow \boldsymbol{\sigma}^{\prime}_{i})$ denotes the change in energy when spin $i$ is flipped in the configuration $\boldsymbol{\sigma}$. Using the Ising Hamiltonian, this energy change can be written as
\begin{equation}
    \Delta \mathcal{H}(\boldsymbol{\sigma}\rightarrow \boldsymbol{\sigma}^{\prime}_{i})\equiv\mathcal{H}(\boldsymbol{\sigma}^{\prime}_{i})-\mathcal{H}(\boldsymbol{\sigma})=2\sigma_{i}\left(J\sum\nolimits_{\langle ij \rangle}\sigma_{j}+H\right).
    \label{dH}
\end{equation}
From the master equation \eqref{masterequation} one directly obtains evolution equations for the first two spin moments (see also Eqs.~(28) and (29) in \cite{Sglauber_timedependent_1963}):
\begin{align}
   \tau\frac{{\rm d}\langle \sigma_{i} \rangle(t)}{{\rm d}t}+\langle \sigma_{i} \rangle(t)  
   &=\langle \sigma_{i}\tanh{(\beta \Delta \mathcal{H}(\boldsymbol{\sigma}\rightarrow \boldsymbol{\sigma}^{\prime}_{i})/2)}\rangle(t), \label{SGlaub1}, \\
   \tau\frac{{\rm d}\langle \sigma_{i}\sigma_{j} \rangle(t)}{{\rm d}t}+2\langle \sigma_{i}\sigma_{j} \rangle(t)
   &=\langle \sigma_{i}\sigma_{j}[\tanh{(\beta \Delta \mathcal{H}(\boldsymbol{\sigma}\rightarrow \boldsymbol{\sigma}^{\prime}_{i})/2)}+\tanh{(\beta \Delta \mathcal{H}(\boldsymbol{\sigma}\rightarrow \boldsymbol{\sigma}^{\prime}_{j})/2)}]\rangle(t). \label{SGlaub2}
\end{align}
Here, the averaging brackets denote averaging w.r.t.~$P(\boldsymbol{\sigma};t)$, i.e., 
\begin{equation}
\langle f \rangle(t) \equiv
\sum_{\boldsymbol{\sigma}}P(\boldsymbol{\sigma};t)f(\boldsymbol{\sigma}),
\end{equation}
for any function $f(\boldsymbol{\sigma})$.
Equations~\eqref{SGlaub1} and \eqref{SGlaub2} are exact, but they are
\emph{not} yet closed because their right-hand sides still involve
higher-order correlations. They provide the starting point for the
derivation of the macroscopic dynamics of $\boldsymbol{x}=(m,s,q)$.

We now derive closed equations for the order parameters within the BG approximation. As in the main text, we partition the lattice into two interpenetrating sublattices $\Lambda^a$ and $\Lambda^b$. Summing Eq.~\eqref{SGlaub1} over all spins on a given sublattice yields
\begin{align}
\begin{split}
    \tau\frac{{\rm d}m^{a}(t)}{{\rm d}t}+m^{a}(t)
   &=\frac{1}{|\Lambda^{a}|}\sum_{i\in\Lambda^a}\langle \sigma_{i}\tanh{(\beta \Delta \mathcal{H}(\boldsymbol{\sigma}\rightarrow \boldsymbol{\sigma}^{\prime}_{i})/2)}\rangle(t),\\
   \tau\frac{{\rm d}m^{b}(t)}{{\rm d}t}+m^{b}(t)
   &=\frac{1}{|\Lambda^{b}|}\sum_{i\in\Lambda^b}\langle \sigma_{i}\tanh{(\beta \Delta \mathcal{H}(\boldsymbol{\sigma}\rightarrow \boldsymbol{\sigma}^{\prime}_{i})/2)}\rangle(t).
\end{split}
\label{BGint}
\end{align}
The task is therefore to evaluate the averages on the right-hand side. A useful simplification follows from Eq.~\eqref{dH}. Since $\tanh(\sigma_i x)=\sigma_i\tanh(x)$ for $\sigma_i=\pm1$, we obtain
\begin{equation}
    \langle \sigma_{i}\tanh{(\beta \Delta \mathcal{H}(\boldsymbol{\sigma}\rightarrow \boldsymbol{\sigma}^{\prime}_{i})/2)}\rangle{\stackrel{\eqref{dH}}{=}}\langle \tanh{(\beta [J\sum\nolimits_{\langle ij \rangle}\sigma_{j}+H])}\rangle.
    \label{trick1}
\end{equation}
Thus, the right-hand side depends only on the local environment of site $i$, i.e., on the number of up spins among its $z$ nearest neighbors. Because the sum $\sum_{\langle ij\rangle}\sigma_j$ can only take the discrete values $-z,-z+2,\dots,z-2,z$, we write
\begin{equation}
    \beta J\sum\nolimits_{\langle ij \rangle}\sigma_{j}+\beta H=(2l-z)\beta J +\beta H,
    \label{dE}
\end{equation}
where $l\in\{0,...,z\}$ denotes the number of neighboring up spins. The expectation value in Eq.~\eqref{trick1} can therefore be decomposed into contributions from the different values of $l$. This gives
\begin{align}
\begin{split}
    \frac{1}{|\Lambda^{a}|}\sum_{i\in\Lambda^a}\langle \sigma_{i}\tanh{(\beta \Delta \mathcal{H}(\boldsymbol{\sigma}\rightarrow \boldsymbol{\sigma}^{\prime}_{i})/2)}\rangle
    &=\sum_{l=0}^{z}\mathcal{P}^{a}_{l}(\boldsymbol{x}_t)\tanh{([2l-z]\beta J+\beta H)}, \\
    \frac{1}{|\Lambda^{b}|}\sum_{i\in\Lambda^b}\langle \sigma_{i}\tanh{(\beta \Delta \mathcal{H}(\boldsymbol{\sigma}\rightarrow \boldsymbol{\sigma}^{\prime}_{i})/2)}\rangle
    &=\sum_{l=0}^{z}\mathcal{P}^{b}_{l}(\boldsymbol{x}_t)\tanh{([2l-z]\beta J+\beta H)},
\end{split}
\end{align}
where $\mathcal{P}^{a}_{l}(\boldsymbol{x})$ and
$\mathcal{P}^{b}_{l}(\boldsymbol{x})$ denote the probabilities that a
spin on sublattice $a$ or $b$ has exactly $l$ neighboring up spins in
a macrostate characterized by $\boldsymbol{x}=(m,s,q)$. We now
evaluate these probabilities within the BG approximation.

For definiteness, consider $\mathcal{P}^{a}_{l}(\boldsymbol{x})$. We first split it according to whether the central spin is up or down:
\begin{equation}
\mathcal{P}^{a}_{l}(\boldsymbol{x})=\mathcal{P}^{a+}_{l}(\boldsymbol{x})+\mathcal{P}^{a-}_{l}(\boldsymbol{x}).
\end{equation}
To obtain explicit expressions, we introduce the numbers of nearest-neighbor pairs of the various types:
\begin{align*}
    N^{ab}_{\pm\pm}&={\rm total \ number \ of \ (\pm,\pm) \ n.n.\ pairs}, \\
    N^{ab}_{\pm\mp}&={\rm total \ number \ of \ (\pm,\mp) \ n.n.\ pairs}.
\end{align*}
These quantities are extensive, i.e., they scale with the total system size $N$. The BG approximation now assumes that nearest-neighbor pairs are randomly distributed throughout the lattice, subject only to the fixed macroscopic values of the relevant densities. Under this assumption,
\begin{align}
\begin{split}
\mathcal{P}^{a+}_{l}(\boldsymbol{x})&=\frac{N^{a}_{+}}{N/2}\times 
\binom{N^{ab}_{++}}{l}\binom{N^{ab}_{+-}}{z-l}/\binom{N^{ab}_{++}+N^{ab}_{+-}}{z}, \\
\mathcal{P}^{a-}_{l}(\boldsymbol{x})&=\frac{N^{a}_{-}}{N/2}\times 
  \binom{N^{ab}_{-+}}{l}\binom{N^{ab}_{--}}{z-l}/\binom{N^{ab}_{-+}+N^{ab}_{--}}{z},
\end{split}
\label{SBG1}
\end{align}
where we used $|\Lambda^{a}|=|\Lambda^{b}|=N/2$, and $N^{a}_{\pm}$ denotes the total number of up/down spins on sublattice $\Lambda^{a}$, given by
\begin{equation}
    z N^{a}_{\pm}=N^{ab}_{\pm+}+N^{ab}_{\pm-}.
\end{equation}
To rewrite everything in terms of the order parameters, we next express the pair numbers through $(m^{a},m^{b},q)$. By definition,
\begin{equation}
    q\equiv \frac{2}{zN}\sum_{\langle ij \rangle} \langle \sigma_i \sigma_j\rangle=\frac{2}{zN}(N^{ab}_{++}+N^{ab}_{--}-N^{ab}_{+-}-N^{ab}_{-+}),
    \label{q}
\end{equation}
where the factor of $2/zN$ on the right-hand side represents the total number of pairs on a lattice with coordination number $z$, i.e. 
\begin{equation}
    zN/2=N^{ab}_{++}+N^{ab}_{--}+N^{ab}_{+-}+N^{ab}_{-+}.
\end{equation}
Likewise, the magnetizations on the two sublattices can be written as
\begin{align}
\begin{split}
    m^{a}&=\frac{2}{zN}\left(N^{ab}_{++}+N^{ab}_{+-}-N^{ab}_{-+}-N^{ab}_{--}\right), \\
    m^{b}&=\frac{2}{zN}\left(N^{ab}_{++}+N^{ab}_{-+}-N^{ab}_{+-}-N^{ab}_{--}\right).
\end{split}
\label{SLmag}
\end{align}
Together, Eqs.~\eqref{q}--\eqref{SLmag} form four linear equations for the four unknowns $\{N^{ab}_{++},N^{ab}_{+-},N^{ab}_{-+},N^{ab}_{--}\}$. Solving them gives
\begin{align}
\begin{split}
    N^{ab}_{\pm\pm}&=(zN/8)(1\pm m^a\pm m^b+q),\\
    N^{ab}_{\pm\mp}&=(zN/8)(1\pm m^a\mp m^b-q).
\end{split}
\label{pairrel}
\end{align}
Substituting Eqs.~\eqref{pairrel} into Eq.~\eqref{SBG1} gives $\mathcal{P}^{a\pm}_{l}(\boldsymbol{x})$ in closed form in terms of $(m^a,m^b,q)$, and therefore also in terms of $\boldsymbol{x}=(m,s,q)$ through the relations
\begin{align}
\begin{split}
    m&\equiv (m^a+m^b)/2,\\
    s&\equiv (m^a-m^b)/2.
\end{split}
\end{align}
The final simplification is obtained in the thermodynamic limit
$N\to\infty$, where we use the following asymptotic result for the binomial coefficient
\begin{equation}
    \lim\limits_{N\rightarrow\infty}\binom{Nx}{l}= \frac{(Nx)^{l}}{l!}+\mathcal{O}(N^{l-1}), \ l \in \mathbb{N}.
\end{equation}
This in turn yields
\begin{align}
\begin{split}
\lim\limits_{N\rightarrow\infty}\mathcal{P}^{a+}_{l}(\boldsymbol{x})&\simeq \binom{z}{l}\frac{(1+m^{a})^{1-z}}{2^{z+1}}\frac{(1+m^{a}+m^{b}+q)^{l}}{(1+m^{a}-m^{b}-q)^{l-z}},\\
\lim\limits_{N\rightarrow\infty}\mathcal{P}^{a-}_{l}(\boldsymbol{x})&\simeq\binom{z}{l}\frac{(1-m^{a})^{1-z}}{2^{z+1}}\frac{(1-m^{a}+m^{b}-q)^{l}}{(1-m^{a}-m^{b}+q)^{l-z}}.
\end{split}
\label{PbgS}
\end{align}
The corresponding expressions for $\mathcal{P}^{b\pm}_{l}(\boldsymbol{x})$ follow by exchanging $a\leftrightarrow b$. Since $q$ appears explicitly in Eqs.~\eqref{PbgS}, we also need an equation for its time evolution. This can be obtained from Eq.~\eqref{SGlaub2} in the same way as above. The resulting BG dynamics is the closed system of three coupled autonomous ordinary differential equations
\begin{align}
\begin{split}
    \tau\frac{{\rm d} m^{a}(t)}{{\rm d}t}+m^{a}(t)&=\sum_{l=0}^{z}\mathcal{P}^{a}_{l}(\boldsymbol{x}_t)\tanh{([2l-z]\beta J+\beta H)},\\
    \tau\frac{{\rm d} m^{b}(t)}{{\rm d}t}+m^{b}(t)&=\sum_{l=0}^{z}\mathcal{P}^{b}_{l}(\boldsymbol{x}_t)\tanh{([2l-z]\beta J+\beta H)},\\
    \tau\frac{{\rm d} q(t)}{{\rm d}t}+2q(t)&=\sum_{l=0}^{z}\alpha_l[\mathcal{P}^{a}_{l}(\boldsymbol{x}_t)+\mathcal{P}^{b}_{l}(\boldsymbol{x}_t)]\tanh{([2l-z]\beta J+\beta H)},
\end{split}
\end{align}
where $\alpha_l\equiv 2l/z-1$. To recover the compact form used in the main text, it is convenient to introduce the coarse-grained transition rates
\begin{equation}\label{BGrates}
w_{l}^{\mu \pm}(\boldsymbol{x};T) \equiv \mathcal{P}^{\mu \pm}_{l}(\boldsymbol{x})[1\mp \tanh([2l-z]\tilde{J}+ \tilde{H})]/2\tau,
\qquad \mu\in\{a,b\},
\end{equation}
with $\tilde J=\beta J$ and  $\tilde H=\beta H$. The probabilities $\mathcal{P}^{\mu\pm}_l$ satisfy the sum rules
\begin{equation}
\sum_{l=0}^{z}[\mathcal{P}^{\mu+}_l(\boldsymbol{x})-\mathcal{P}^{\mu-}_l(\boldsymbol{x})]=m^\mu,
\qquad
\sum_{l=0}^{z}\alpha_l[\mathcal{P}^{\mu+}_l(\boldsymbol{x})-\mathcal{P}^{\mu-}_l(\boldsymbol{x})]=q.
\label{sumrules}
\end{equation}
Using these identities, the BG equations can be rewritten as
\begin{align}
\begin{split}
\tau\frac{{\rm d} m^{\mu}(t)}{{\rm d}t}
&=
2\sum_{l=0}^{z} \left[w_{l}^{\mu -}(\boldsymbol{x}_t;T)-w_{l}^{\mu +}(\boldsymbol{x}_t;T)\right],\\
\tau\frac{{\rm d} q(t)}{{\rm d}t}
&=
2\sum_{\mu}\sum_{l=0}^{z}
\alpha_l\left[w_{l}^{\mu -}(\boldsymbol{x}_t;T)-w_{l}^{\mu +}(\boldsymbol{x}_t;T)\right],
\end{split}
\label{SBGeqs}
\end{align}
which is equivalent to Eq.~(7) in the main manuscript. 
%---------------------------------
%---------------------------------
%---------------------------------
\section{Derivation of the free energy}
%---------------------------------
%---------------------------------
%---------------------------------
\noindent Having derived the probabilities
$\mathcal{P}^{a\pm}_{l}(\boldsymbol{x})$ and
$\mathcal{P}^{b\pm}_{l}(\boldsymbol{x})$, we can now use them to
construct the BG free energy. Since the previous subsection already
expressed these probabilities in terms of the macroscopic order
parameters, we here only summarize the ingredients needed for the free-energy functional. The quantities $\mathcal{P}^{\mu\pm}_{l}$ are joint probabilities: they give the probability that a central spin on sublattice $\mu\in\{a,b\}$ has orientation $\sigma=\pm1$ and exactly $l$ neighboring up spins. By construction, they satisfy the normalization condition
\begin{equation}
\sum_{l=0}^{z}[\mathcal P_l^{\mu+}(\boldsymbol{x})+\mathcal P_l^{\mu-}(\boldsymbol{x})]=1.
\end{equation}
Summing over $l$ gives the corresponding single-site probabilities,
\begin{equation}
P_\pm^\mu=\sum_{l=0}^{z}\mathcal P_l^{\mu\pm}(\boldsymbol{x})=\frac{1\pm m^{\mu}}{2}.
\end{equation}
The nearest-neighbor pair probabilities are obtained from the average number of up and down neighbors surrounding a central spin. For instance,
\begin{equation}
P_{++}^{ab}(\boldsymbol{x})
=
\frac{1}{z}\sum_{l=0}^{z}l \, \mathcal P_l^{a+}(\boldsymbol{x}),
\qquad
P_{+-}^{ab}(\boldsymbol{x})
=
\frac{1}{z}\sum_{l=0}^{z}(z-l)\,\mathcal P_l^{a+}(\boldsymbol{x}),
\end{equation}
and analogous relations hold for $P_{--}^{ab}$ and $P_{-+}^{ab}$. Evaluating these sums using Eq.~\eqref{PbgS} gives
\begin{align}
\begin{split}
P^{ab}_{\pm\pm}(\boldsymbol{x})&=(1\pm m^a \pm m^b+q)/4, \\
P^{ab}_{\pm\mp}(\boldsymbol{x})&=(1\pm m^a\mp m^b-q)/4.
\end{split}
\label{Ppair}
\end{align}
The BG entropy follows from the cluster-variational construction
\cite{SPhysRev.81.988}. In this approximation, the entropy is built
from nearest-neighbor pair marginals and single-site marginals. The
contribution from all nearest-neighbor pairs is counted once, however,
this overcounts the site contribution: each site belongs to $z$ different nearest-neighbor pairs, whereas it should contribute only once to the total entropy. One therefore has to subtract the excess $z-1$ site contributions. This gives
\begin{equation}
S_{\rm BG}(\boldsymbol{x})
=
S_{\rm pair}(\boldsymbol{x})-(z-1)S_{\rm site}(\boldsymbol{x}),
\end{equation}
where
\begin{align}
S_{\rm pair}(\boldsymbol{x})
&=
-\frac{zN}{2}
\sum_{\sigma=\pm}\sum_{\sigma'=\pm}
P^{ab}_{\sigma\sigma'}\ln(P^{ab}_{\sigma\sigma'}),\\
S_{\rm site}(\boldsymbol{x})
&=
-\frac{N}{2}
\sum_{\mu}\sum_{\sigma=\pm}
P^{\mu}_\sigma\ln(P^{\mu}_\sigma).
\end{align}
Here the prefactor $zN/2$ is the total number of nearest-neighbor pairs in a lattice with coordination number $z$. Substituting the site and pair probabilities above, and subsequently dividing by $N$, yields an explicit expression for the BG entropy. To obtain the BG free energy, we combine this entropy with the average energy. Using the definitions of the order parameters, one finds
\begin{equation}
\langle \mathcal{H}(\boldsymbol{\sigma}) \rangle
=
-J\sum_{\langle ij\rangle}\langle \sigma_{i}\sigma_{j}\rangle
-
H \sum_{i} \langle \sigma_{i} \rangle
=
N\left[-zJq/2-Hm\right].
\end{equation}
The BG free energy per spin therefore reads
\begin{equation}
\mathcal F(\boldsymbol{x};T)
=
\frac{1}{N}\left[\langle \mathcal H(\boldsymbol{\sigma})  \rangle
-
\beta^{-1} S_{\rm BG}(\boldsymbol{x})\right],
\end{equation}
which coincides with Eq.~(8) in the main manuscript where $\mathcal{S}_{\rm BG}(\boldsymbol{x})\equiv S_{\rm BG}(\boldsymbol{x})/N$.
%---------------------------------
%---------------------------------
%---------------------------------
\section{Thermomajorization of the finite-$N$ BG dynamics}
%---------------------------------
%---------------------------------
%---------------------------------
\noindent The strong Mpemba effect discussed in the main text arises because the initial paramagnetic state has zero overlap with the slowest relaxation mode. It is therefore expected to be independent of the particular distance measure used to quantify relaxation. Here we make this statement precise using the thermomajorization criterion of Ref.~\cite{SPhysRevLett.134.107101SM}.
%---------------------------------
%---------------------------------
\subsection{Finite-$N$ BG dynamics}
%---------------------------------
%---------------------------------
\noindent The BG equations~\eqref{SBGeqs} describe the deterministic evolution of the macroscopic variables in the limit $N\to\infty$. To apply the finite-state thermomajorization criterion, we introduce the corresponding finite-$N$ stochastic BG dynamics. Its state space $\Omega_N$ consists of the admissible discrete values of $\boldsymbol{x}=(m^a,m^b,q)$. We denote the probability vector over this state space by $\boldsymbol{P}_N(t)$ and its component associated with the macrostate $\boldsymbol{x}$ by $P_N(\boldsymbol{x},t)$. Its evolution is governed by
\begin{equation}
    \frac{{\rm d}\boldsymbol{P}_N(t)}{{\rm d}t}
    =
    W_N\boldsymbol{P}_N(t).
    \label{eq:tm-master}
\end{equation}
The off-diagonal elements of the generator are
\begin{equation}
    \bigl(W_N\bigr)_{\boldsymbol{x}^{\mu\pm}_l,\boldsymbol{x}}
    =
    (N/2)
    w_{l,N}^{\mu\pm}(\boldsymbol{x};T_{\rm f}),
    \label{eq:tm-Woff}
\end{equation}
where $\boldsymbol{x}^{\mu\pm}_l$ is the macrostate obtained by flipping a spin on sublattice $\mu\in\{a,b\}$ with $l$ up-spin neighbors. The rates $w_{l,N}^{\mu\pm}$ are evaluated using the finite-$N$ neighbor statistics given by Eq.~\eqref{SBG1}. Probability conservation fixes the diagonal elements as
\begin{equation}
    \bigl(W_N\bigr)_{\boldsymbol{x},\boldsymbol{x}}
    =
    -(N/2)
    \sum_{\mu\in\{a,b\}}\sum_{l=0}^{z}
    [
        w_{l,N}^{\mu+}(\boldsymbol{x};T_{\rm f})
        +
        w_{l,N}^{\mu-}(\boldsymbol{x};T_{\rm f})
   ].
    \label{eq:tm-Wdiag}
\end{equation}
Each transition changes the macroscopic variables by $\mathcal{O}(N^{-1})$ and occurs at a total rate of $\mathcal{O}(N)$. Consequently, the fluctuations vanish as $N^{-1/2}$, while the mean drift converges to Eq.~\eqref{SBGeqs}. The law-of-large-numbers limit of Eq.~\eqref{eq:tm-master} therefore recovers the deterministic BG dynamics. We denote the stationary probability vector of $W_N$ by $\boldsymbol{\pi}_N$, with components $\pi_N(\boldsymbol{x})$. The generator satisfies detailed balance,
\begin{equation}
    \bigl(W_N\bigr)_{\boldsymbol{x}',\boldsymbol{x}}
    \pi_N(\boldsymbol{x})
    =
    \bigl(W_N\bigr)_{\boldsymbol{x},\boldsymbol{x}'}
    \pi_N(\boldsymbol{x}'),
    \label{eq:tm-detailed-balance}
\end{equation}
for every pair of connected macrostates $\boldsymbol{x}$ and $\boldsymbol{x}'$. Consequently,
$\boldsymbol{\Pi}_N^{-1/2}W_N\boldsymbol{\Pi}_N^{1/2}$, with
$\boldsymbol{\Pi}_N=\operatorname{diag}(\boldsymbol{\pi}_N)$, is symmetric. The eigenvalues of $W_N$ are therefore real and can be ordered as
\begin{equation}
    0=\lambda_{1,N}>
    \lambda_{2,N}>
    \lambda_{3,N}\geq\cdots.
\end{equation}
Assuming that the slowest nonzero eigenvalue is nondegenerate, the relaxation from an initial distribution $\boldsymbol{P}_{N}^{\alpha}(0)$ can be written as
\begin{equation}
    \boldsymbol{P}_N^{\alpha}(t)
    =
    \boldsymbol{\pi}_N
    +
    \sum_{n\geq2}
    a_{n,N}^{\alpha}
    {\rm e}^{\lambda_{n,N}t}
    \boldsymbol{r}_{n,N},
    \label{eq:tm-spectral}
\end{equation}
where $\boldsymbol{r}_{n,N}$ and $\boldsymbol{\ell}_{n,N}$ are the $n$th right and left eigenvectors of $W_N$, respectively. Choosing them biorthonormal, the amplitudes are
\begin{equation}
    a_{n,N}^{\alpha}
    =
    \boldsymbol{\ell}_{n,N}^{\rm T}
    \left(
        \boldsymbol{P}_{N}^{\alpha}(0)
        -
        \boldsymbol{\pi}_N
    \right).
    \label{eq:tm-amplitude}
\end{equation}
To represent the initial conditions considered in the main text, we take $\boldsymbol{P}_{N}^{\rm P}(0)$ to be concentrated around the paramagnetic equilibrium macrostate and $\boldsymbol{P}_{N}^{\rm AF,+}(0)$ to be concentrated around the symmetry-broken antiferromagnetic branch with $s>0$.
%---------------------------------
%---------------------------------
\subsection{Thermomajorization}
%---------------------------------
%---------------------------------
\noindent Consider two probability vectors $\boldsymbol{p}$ and $\boldsymbol{q}$ relaxing toward the same stationary distribution $\boldsymbol{\pi}$. We write
$\boldsymbol{p}\prec_{\boldsymbol{\pi}}\boldsymbol{q}$ when $\boldsymbol{p}$ is thermomajorized by $\boldsymbol{q}$ relative to $\boldsymbol{\pi}$. This means that $\boldsymbol{p}$ is closer to $\boldsymbol{\pi}$ than $\boldsymbol{q}$ according to every measure that decreases monotonically under the Markov dynamics. The thermomajorization relation is equivalent to
\begin{equation}
    \left\|
        \boldsymbol{p}-z\boldsymbol{\pi}
    \right\|_1
    \leq
    \left\|
        \boldsymbol{q}-z\boldsymbol{\pi}
    \right\|_1
    \qquad
    \text{for all }z\in\mathbb{R},
    \label{eq:tm-order}
\end{equation}
where $\|\boldsymbol{u}\|_1=\sum_{\boldsymbol{x}}|u(\boldsymbol{x})|$. A thermomajorization Mpemba effect occurs when the initially farther state is thermomajorized by the initially closer state at sufficiently long times.
%---------------------------------
%---------------------------------
\subsection{Sublattice-exchange parity}
\begin{figure*}[t]
    \centering
    \includegraphics[width=0.98\linewidth]{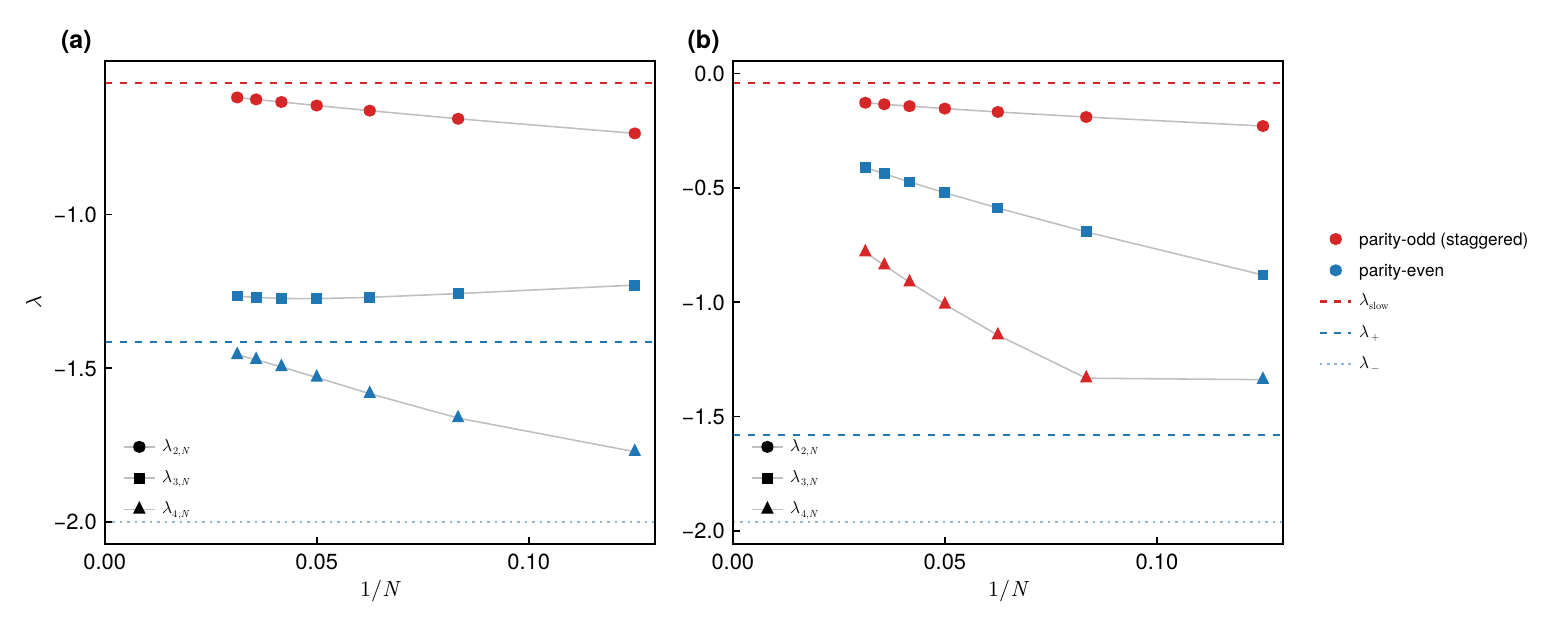}
    \caption{\textbf{Finite-size spectrum and sublattice-exchange parity.}
    The three slowest nonzero eigenvalues $\lambda_{2,N}$, $\lambda_{3,N}$, and $\lambda_{4,N}$ of the finite-$N$ stochastic BG generator $W_N$, plotted as functions of $1/N$ for $N\in\{8,12,16,...,32\}$. Eigenmodes are classified according to their parity under sublattice exchange [see Eq.~\eqref{parity}]: odd (staggered) modes are shown in red and even modes in blue. Results are shown for (a) the cooling protocol with $T_{\rm f}=0.12$ and (b) the heating protocol with $T_{\rm f}=3$. In both cases, the slowest nonstationary mode is odd for every investigated value of $N$. The dashed lines indicate the analytical BG relaxation rates in the paramagnetic phase given by Eqs.~(D6) and~(D9).}
    \label{fig:tm-eigconv}
\end{figure*}

%---------------------------------
%---------------------------------
\noindent Let $\mathcal{R}$ denote the exchange of the two sublattices. Its action on a probability vector $\boldsymbol{p}$ is defined by
\begin{equation}
    [\mathcal{R}\boldsymbol{p}](m,s,q)
    =
   p(m,-s,q),
    \label{eq:tm-parity-operator}
\end{equation}
which is equivalent to exchanging $m^a$ and $m^b$. Since the final equilibrium state is paramagnetic, the generator is invariant under this transformation,
\begin{equation}
    [W_N,\mathcal{R}]=0.
\end{equation}
The left and right eigenvectors of $W_N$ can therefore be chosen to be either even or odd under $\mathcal{R}$. In terms of their components, this means
\begin{equation}
    r_{n,N}(m,-s,q)
    =
    \eta_n r_{n,N}(m,s,q),
    \qquad
    \ell_{n,N}(m,-s,q)
    =
    \eta_n \ell_{n,N}(m,s,q),
    \label{parity}
\end{equation}
where $\eta_n=+1$ for an even mode and $\eta_n=-1$ for an odd mode. As shown in Fig.~\ref{fig:tm-eigconv}, the slowest nonstationary mode is odd for all investigated system sizes $N\in\{8,12,16,...,32\}$, for both the cooling and heating protocols. The paramagnetic initial distribution and the stationary distribution are even under $s\to-s$. Their difference is therefore also even, and its overlap with the odd left eigenvector $\boldsymbol{\ell}_{2,N}$ vanishes exactly:
\begin{equation}
    a_{2,N}^{\rm P}=0.
    \label{eq:tm-zero}
\end{equation}
By contrast, the symmetry-broken antiferromagnetic initial distribution is not even under $s\to-s$ and has, in general, a finite overlap with this mode,
\begin{equation}
    a_{2,N}^{\rm AF,+}\neq0.
    \label{eq:tm-nonzero}
\end{equation}
%---------------------------------
%---------------------------------
\subsection{Long-time thermomajorization}
%---------------------------------
%---------------------------------
\noindent We now apply the long-time thermomajorization criterion of Ref.~\cite{SPhysRevLett.134.107101SM}. Define the contributions of the slowest mode by
\begin{equation}
    \boldsymbol{v}_N^{\rm P}
    =
    a_{2,N}^{\rm P}\boldsymbol{r}_{2,N}
    =
    \boldsymbol{0},
    \qquad
    \boldsymbol{v}_N^{\rm AF,+}
    =
    a_{2,N}^{\rm AF,+}\boldsymbol{r}_{2,N}.
    \label{projections}
\end{equation}
According to this criterion, the paramagnetic distribution is thermomajorized by the antiferromagnetic distribution at sufficiently long times if
\begin{equation}
    \left\|
        \boldsymbol{v}_N^{\rm P}
        -
        z\boldsymbol{\pi}_N
    \right\|_1
    \leq
    \left\|
        \boldsymbol{v}_N^{\rm AF,+}
        -
        z\boldsymbol{\pi}_N
    \right\|_1
    \qquad
    \text{for all }z\in\mathbb{R}.
    \label{eq:tm-longcriterion}
\end{equation}
Since every nonstationary right eigenvector conserves probability,
$\sum_{\boldsymbol{x}}v_N^{\rm AF,+}(\boldsymbol{x})=0$. Therefore,
\begin{align}
    \left\|
        \boldsymbol{v}_N^{\rm P}
        -
        z\boldsymbol{\pi}_N
    \right\|_1
    =
    |z|
    =
    \left|
        \sum_{\boldsymbol{x}}
        \left[
            v_N^{\rm AF,+}(\boldsymbol{x})
            -
            z\pi_N(\boldsymbol{x})
        \right]
    \right|
    \leq
    \left\|
        \boldsymbol{v}_N^{\rm AF,+}
        -
        z\boldsymbol{\pi}_N
    \right\|_1,
    \label{eq:tm-proof}
\end{align}
where the last inequality follows from the triangle inequality. Moreover, the inequality is strict for $z=0$ because $\boldsymbol{v}_N^{\rm AF,+}\neq\boldsymbol{0}$. It follows that, at sufficiently long times,
\begin{equation}
    \boldsymbol{P}_N^{\rm P}(t)
    \prec_{\boldsymbol{\pi}_N}
    \boldsymbol{P}_N^{\rm AF,+}(t).
    \label{eq:tm-lateorder}
\end{equation}
Thus, the paramagnetic initial state eventually becomes closer to the final equilibrium distribution than the antiferromagnetic initial state according to every measure that is monotone under the Markov dynamics. This establishes the observable-independent late-time ordering underlying the strong Mpemba and inverse Mpemba effects for all investigated system sizes $N=8,\ldots,32$.

%---------------------------------
%---------------------------------
%---------------------------------
\section{Relation between the excess free energy and the Kullback-Leibler divergence}
%---------------------------------
%---------------------------------
%---------------------------------
\noindent Here we show that the excess free energy used to quantify relaxation in the main text is directly related to the Kullback-Leibler divergence. Let $\boldsymbol{P}_N(t)$ denote the time-dependent probability vector over the discrete macrostate space $\Omega_N$ for a system containing $N$ spins. Its component $P_N(\boldsymbol{x},t)$ is the probability of finding the system in the macrostate $\boldsymbol{x}=(m,s,q)$ at time $t$. In the thermodynamic limit, these probabilities have the large-deviation form \cite{SRevModPhys.97.015002}
\begin{equation}
P_N(\boldsymbol{x},t)
\asymp
\exp[-N\mathcal V(\boldsymbol{x},t)].
\label{LDPtime}
\end{equation}
Here, $\asymp$ denotes logarithmic asymptotic equivalence in the limit $N\to\infty$, i.e.,
\begin{equation}
\lim_{N\to\infty}
\frac{1}{N}\ln P_N(\boldsymbol{x},t)
=
-\mathcal V(\boldsymbol{x},t).
\end{equation}
Subexponential contributions to $P_N(\boldsymbol{x},t)$ are therefore not included in Eq.~\eqref{LDPtime}. The time-dependent rate function $\mathcal V(\boldsymbol{x},t)$ obeys a Hamilton--Jacobi equation \cite{SRevModPhys.97.015002}. Its minimum follows the deterministic BG dynamics, such that
\begin{equation}
\mathcal V(\boldsymbol{x}_t,t)=0,
\qquad
\nabla_{\boldsymbol{x}}\mathcal V(\boldsymbol{x}_t,t)=0,
\label{Vminimum}
\end{equation}
where $\boldsymbol{x}_t$ is the solution of Eq.~\eqref{SBGeqs}. Let $\boldsymbol{\pi}_N$ denote the stationary probability vector at the final temperature $T_{\rm f}$, with components $\pi_N(\boldsymbol{x})$. These probabilities similarly have the large-deviation form
\begin{equation}
\pi_N(\boldsymbol{x})
\asymp
\exp[-N\mathcal V_{\rm eq}(\boldsymbol{x})],
\label{LDPeq}
\end{equation}
where the equilibrium rate function is
\begin{equation}
\mathcal V_{\rm eq}(\boldsymbol{x})
=
\beta_{\rm f}
\left[
\mathcal F(\boldsymbol{x};T_{\rm f})
-
\mathcal F(\boldsymbol{x}_{\rm eq}(T_{\rm f});T_{\rm f})
\right].
\label{Veq}
\end{equation}
The additive constant has been chosen such that $\mathcal V_{\rm eq}(\boldsymbol{x}_{\rm eq}(T_{\rm f}))=0$. The Kullback-Leibler divergence between the complete time-dependent and stationary probability distributions is defined as
\begin{equation}
D_{\rm KL}
\left(
\boldsymbol{P}_N(t)
\middle\|
\boldsymbol{\pi}_N
\right)
\equiv
\sum_{\boldsymbol{x}\in\Omega_N}
P_N(\boldsymbol{x},t)
\ln
\frac{P_N(\boldsymbol{x},t)}
{\pi_N(\boldsymbol{x})}.
\label{DKLdef}
\end{equation}
Using Eqs.~\eqref{LDPtime} and \eqref{LDPeq}, the logarithm in Eq.~\eqref{DKLdef} becomes
\begin{equation}
\ln
\frac{P_N(\boldsymbol{x},t)}
{\pi_N(\boldsymbol{x})}
=
N
\left[
\mathcal V_{\rm eq}(\boldsymbol{x})
-
\mathcal V(\boldsymbol{x},t)
\right]
+
o(N),
\end{equation}
and hence
\begin{align}
D_{\rm KL}
\left(
\boldsymbol{P}_N(t)
\middle\|
\boldsymbol{\pi}_N
\right)
&=
N
\sum_{\boldsymbol{x}\in\Omega_N}
P_N(\boldsymbol{x},t)
\left[
\mathcal V_{\rm eq}(\boldsymbol{x})
-
\mathcal V(\boldsymbol{x},t)
\right]
+
o(N).
\label{DKLLDP}
\end{align}
For $N\to\infty$, $\boldsymbol{P}_N(t)$ becomes concentrated around the minimum $\boldsymbol{x}_t$ of $\mathcal V(\boldsymbol{x},t)$. Assuming that this minimum is unique, the saddle-point approximation together with Eq.~\eqref{Vminimum} gives
\begin{align}
\lim_{N\to\infty}
\frac{1}{N}
D_{\rm KL}
\left(
\boldsymbol{P}_N(t)
\middle\|
\boldsymbol{\pi}_N
\right)
=
\mathcal V_{\rm eq}(\boldsymbol{x}_t)
-
\mathcal V(\boldsymbol{x}_t,t)
=
\mathcal V_{\rm eq}(\boldsymbol{x}_t)
=
\beta_{\rm f}
\left[
\mathcal F(\boldsymbol{x}_t;T_{\rm f})
-
\mathcal F(\boldsymbol{x}_{\rm eq}(T_{\rm f});T_{\rm f})
\right]
=
\beta_{\rm f}D(t),
\label{DKLfreeenergy}
\end{align}
where $D(t)$ is the excess free energy given by Eq.~(11) in the main text. Thus, $\beta_{\rm f}D(t)$ coincides with the Kullback-Leibler divergence per spin between the time-dependent and final equilibrium probability distributions in the thermodynamic limit.
%---------------------------------
%---------------------------------
%---------------------------------
\section{Concavity of the phase boundary and conditions for reentrance}
%---------------------------------
%---------------------------------
%---------------------------------
\noindent In this section, we determine for which values of the coordination number $z$ the phase diagram becomes reentrant. Having derived the free energy and the critical magnetic field $H_c(T)$, we analyze the shape of the phase boundary by studying its second derivative:
\begin{equation}
    \frac{d^2 H_{c}(T)}{dT^2} = -\frac{J^2 (z-2) (z-1) z \operatorname{csch}^4\left(J/T\right) \sqrt{\operatorname{tanh} \left(J/T\right)+z-1}}{2 T^3 \left(\operatorname{coth} \left(J/T\right)+z-1\right)^{3/2} \left((z-1) \operatorname{coth} \left(J/T\right)+1\right)^2},
\end{equation}
where, for simplicity, we have set $k_{\rm B}=1$. We now examine the sign of this expression in the antiferromagnetic case, $J<0$. Since the coordination number satisfies $z>2$, all factors in the numerator are positive. In particular, the square-root factor is well defined because $\tanh(J/T)>-1$, so
\begin{equation}
\tanh\left(J/T\right)+z-1>z-2>0.
\end{equation}
To determine the sign of the denominator, recall that the critical magnetic field is defined only for temperatures in the range
\begin{equation}
0 < T < T_c(H=0) \equiv -\frac{2J}{\ln\left(\frac{z}{z-2}\right)}.
\label{temperaturerange}
\end{equation}
Within this interval, all factors in the denominator are also positive. In particular, we must show that
\begin{equation}
\operatorname{coth}\left(J/T\right)+z-1> 0.
\end{equation}
This inequality can be rewritten as $J/T>\operatorname{arccoth}(1-z)$. Using
$\operatorname{arccoth}(x)=\frac{1}{2}\ln\left(\frac{1+1/x}{1-1/x}\right)$,
we obtain
\begin{equation}
    \frac{J}{T} > -\frac{1}{2}\ln{\left(\frac{z}{z-2}\right)}.
\end{equation}
This is equivalent to Eq.~\eqref{temperaturerange}. Hence the denominator is positive throughout the physically relevant temperature range. Therefore, for the antiferromagnetic case with $J<0$ and $z>2$, we find
\begin{equation}
    \frac{d^2 H_{c}(T)}{dT^2}<0.
\end{equation}
Thus, $H_c(T)$ is concave throughout its full domain of definition, and therefore its derivative $dH_c/dT$ is a strictly decreasing function of temperature. This immediately constrains the possible shape of the phase boundary. If the low-temperature slope is negative, then $dH_c/dT<0$ for all $T$, and the phase boundary is monotonic. By contrast, if the low-temperature slope is positive, then concavity alone implies that the slope must decrease with temperature. Since $H_c(T)$ must vanish at $T=T_c(H=0)$, a positive initial slope is incompatible with monotonic growth over the full interval $0<T<T_c(H=0)$. The derivative must therefore cross zero at a finite temperature, which means that $H_c(T)$ develops a maximum and is non-monotonic. Evaluating the low-temperature behavior of $H_c(T)$ shows that its slope is positive if and only if $z>z^*$. We therefore conclude that the phase boundary is non-monotonic, and hence reentrant, for $z>z^*$, while it is monotonic for $z<z^*$. This is consistent with the behavior discussed in the main manuscript.
%---------------------------------
%---------------------------------
%---------------------------------
\section{Linear stability analysis around equilibrium}
%---------------------------------
%---------------------------------
%---------------------------------
\noindent In this section we analyze the linear stability of Eq.~\eqref{SBGeqs}. We define the currents as
\begin{equation}
J_l^\mu(\boldsymbol{x};T)\equiv w_l^{\mu -}(\boldsymbol{x};T)-w_l^{\mu +}(\boldsymbol{x};T).
\end{equation}
Let $\boldsymbol{x}_{\rm eq}=\boldsymbol{x}_{\rm eq}(T)$ be the equilibrium state of Eq.~\eqref{SBGeqs} at temperature $T$. By definition,
\begin{equation}
w_l^{\mu +}(\boldsymbol{x}_{\rm eq};T)=w_l^{\mu -}(\boldsymbol{x}_{\rm eq};T)\equiv w_{l,{\rm eq}}^\mu ,
\end{equation}
for all $\mu\in\{a,b\}$ and $l\in\{0,\dots,z\}$. We now consider small perturbations $\boldsymbol{\delta x}_t\equiv \boldsymbol{x}_t-\boldsymbol{x}_{\rm eq}$ such that $|\boldsymbol{\delta x}_t|\ll 1$. To linearize the currents around equilibrium, we employ the local detailed-balance relation
\begin{equation}
\ln\biggl(
\frac{w_l^{\mu +}(\boldsymbol{x};T)}{w_l^{\mu -}(\boldsymbol{x};T)}
\biggr)=4[\partial_{m^\mu}\tilde{\mathcal F}(\boldsymbol{x};T)
+
\alpha_l\partial_q\tilde{\mathcal F}(\boldsymbol{x};T)],
\label{cg_db_sup}
\end{equation}
where $\tilde{\mathcal F}(\boldsymbol{x};T)=\beta\mathcal F(\boldsymbol{x};T)$ with $\beta\equiv 1/k_{\rm B}T$, and $\partial_x \tilde{\mathcal F}(\boldsymbol{x};T)\equiv \partial \tilde{\mathcal F}(\boldsymbol{x};T)/\partial x$. Expanding both sides of Eq.~\eqref{cg_db_sup} to linear order in $\boldsymbol{\delta x}_t$ yields
\begin{equation}
\ln\biggl(
\frac{w_l^{\mu +}(\boldsymbol{x}_{\rm eq}+\boldsymbol{\delta x}_t;T)}{w_l^{\mu -}(\boldsymbol{x}_{\rm eq}+\boldsymbol{\delta x}_t;T)}
\biggr)\simeq\frac{1}{w^{\mu}_{l,{\rm
      eq}}}\boldsymbol{\nabla}\left.[w_l^{\mu
    +}(\boldsymbol{x};T)-w_l^{\mu -}(\boldsymbol{x};T)]\right |_{\boldsymbol{x}_{\rm eq}}\cdot \boldsymbol{\delta x}_t=\frac{-1}{w^{\mu}_{l,{\rm eq}}}\left.\boldsymbol{\nabla}J_l^{\mu}(\boldsymbol{x};T)\right|_{\boldsymbol{x}_{\rm eq}}\cdot \boldsymbol{\delta x}_t,
\end{equation}
\begin{equation}
4[\partial_{m^\mu}\tilde{\mathcal F}(\boldsymbol{x}_{\rm eq}+\boldsymbol{\delta x}_t;T)
+
\alpha_l\partial_q\tilde{\mathcal F}(\boldsymbol{x}_{\rm eq}+\boldsymbol{\delta x}_t;T)]\simeq 
4 \boldsymbol{\nabla}\left.[\partial_{m^\mu}\tilde{\mathcal F}(\boldsymbol{x};T)
+
\alpha_l\partial_q\tilde{\mathcal F}(\boldsymbol{x};T)]\right|_{\boldsymbol{x}_{\rm eq}}\cdot \boldsymbol{\delta x}_t,
\end{equation}
where
\begin{equation}
    \boldsymbol{\nabla}\equiv\left(\partial_{m},\partial_{s},\partial _q\right)^{\rm T}.
\end{equation}
Combining these expressions, we obtain the linearized currents in terms of the second derivatives of the free energy:
\begin{equation}
J_l^\mu(\boldsymbol{x}_{\rm eq}+\boldsymbol{\delta x}_t;T)
\simeq \left.\boldsymbol{\nabla}J_l^{\mu}(\boldsymbol{x};T)\right|_{\boldsymbol{x}_{\rm eq}}\cdot \boldsymbol{\delta x}_t=-4w^{\mu}_{l,{\rm eq}} \boldsymbol{\nabla}\left.[\partial_{m^\mu}\tilde{\mathcal F}(\boldsymbol{x};T)
+
\alpha_l\partial_q\tilde{\mathcal F}(\boldsymbol{x};T)]\right|_{\boldsymbol{x}_{\rm eq}}\cdot \boldsymbol{\delta x}_t.
\end{equation}
Inserting this expression into Eq.~\eqref{SBGeqs} and transforming to the $(m,s,q)$ basis, we obtain the closed linear system reported as Eq.~(C1) in the main manuscript.
%---------------------------------
%---------------------------------
%---------------------------------
\section{Long-time relaxation for the inverse Mpemba protocol}
\begin{figure}
    \centering
    \includegraphics[width=0.7\linewidth]{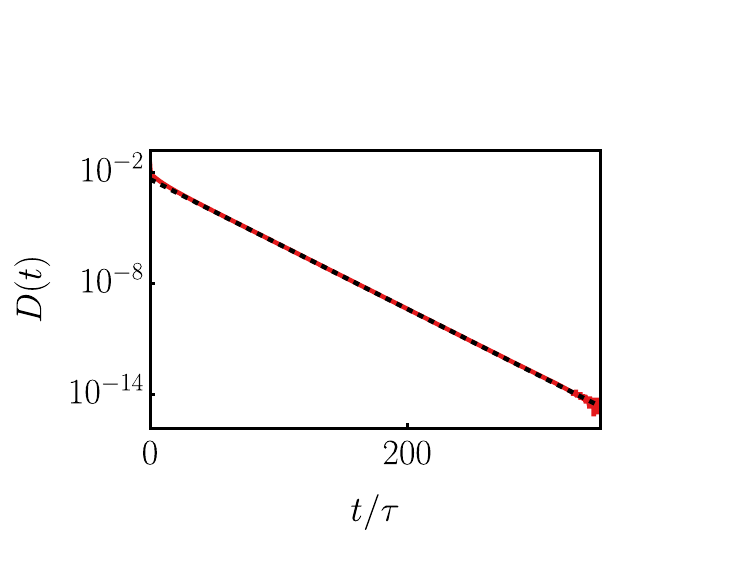}
    \caption{\textbf{Long-time behavior of the inverse Mpemba protocol.}
    Excess free energy $D(t)$ (see Eq.~(11) in the main manuscript) as a function of time for the hotter initial state (red), shown over the extended interval $0\le t/\tau\le 350$. The parameter values are the same as in Fig.~2(b) of the main manuscript. The black dashed line shows the analytical long-time asymptotic expression given by Eq.~(E3) in the main manuscript. The small wiggles visible at very large times are numerical artifacts that arise because the excess free energy become extremely small.}
    \label{FigSM}
\end{figure}

%---------------------------------
%---------------------------------
%---------------------------------
\noindent To make the slow relaxation of the hotter initial state in  Fig.~2(b)  more clearly visible, we show in Fig.~\ref{FigSM} the same figure over a much longer time window, extending to $t/\tau=350$. In the main figure, the much steeper initial decay of the colder state makes the relaxation of the hotter state appear almost flat on the shown scale. The extended-time plot demonstrates that the hotter state indeed continues to relax and follows the expected asymptotic behavior.
%---------------------------------
%---------------------------------
%---------------------------------
%

\end{document}